\documentclass[a4paper,twoside,12pt]{article}
\usepackage{amsmath}
\usepackage{amsfonts}
\usepackage{amssymb}
\usepackage{amsthm}
\usepackage{amscd}

\input xy
\xyoption{all} 

  \usepackage{paralist}
  \usepackage{graphics} 
  \usepackage{epsfig} 
\usepackage{graphicx}  
 \usepackage[colorlinks=true]{hyperref}
   \hypersetup{urlcolor=blue, citecolor=red}

\usepackage{floatrow}
\usepackage{latexsym}

\newfloat{pict}{h}{lob}
\floatname{pict}{Fig.}

\font\black=cmbx10 \font\sblack=cmbx7 \font\ssblack=cmbx5 \font\blackital=cmmib10  \skewchar\blackital='177
\font\sblackital=cmmib7 \skewchar\sblackital='177 \font\ssblackital=cmmib5 \skewchar\ssblackital='177
\font\sanss=cmss10 \font\ssanss=cmss8 
\font\sssanss=cmss8 scaled 600 \font\blackboard=msbm10 \font\sblackboard=msbm7 \font\ssblackboard=msbm5
\font\caligr=eusm10 \font\scaligr=eusm7 \font\sscaligr=eusm5  \font\fraktur=eufm10
\font\sfraktur=eufm7 \font\ssfraktur=eufm5

\font\bsymb=cmsy10 scaled\magstep2
\def\all#1{\setbox0=\hbox{\lower1.5pt\hbox{\bsymb
       \char"38}}\setbox1=\hbox{$_{#1}$} \box0\lower2pt\box1\;}
\def\exi#1{\setbox0=\hbox{\lower1.5pt\hbox{\bsymb \char"39}}
       \setbox1=\hbox{$_{#1}$} \box0\lower2pt\box1\;}

\newfam\bifam
\textfont\bifam=\blackital \scriptfont\bifam=\sblackital \scriptscriptfont\bifam=\ssblackital

\newfam\blfam
\textfont\blfam=\black \scriptfont\blfam=\sblack \scriptscriptfont\blfam=\ssblack

\newfam\bbfam
\textfont\bbfam=\blackboard \scriptfont\bbfam=\sblackboard \scriptscriptfont\bbfam=\ssblackboard

\newfam\ssfam
\textfont\ssfam=\sanss \scriptfont\ssfam=\ssanss \scriptscriptfont\ssfam=\sssanss
\def\sss#1{{\fam\ssfam\relax#1}}

\newfam\clfam
\textfont\clfam=\caligr \scriptfont\clfam=\scaligr \scriptscriptfont\clfam=\sscaligr

\newfam\frfam
\textfont\frfam=\fraktur \scriptfont\frfam=\sfraktur \scriptscriptfont\frfam=\ssfraktur

\def\hpb#1{\setbox0=\hbox{${#1}$}
    \copy0 \kern-\wd0 \kern.2pt \box0}
\def\vpb#1{\setbox0=\hbox{${#1}$}
    \copy0 \kern-\wd0 \raise.08pt \box0}

\def\pmb#1{\setbox0\hbox{${#1}$} \copy0 \kern-\wd0 \kern.2pt \box0}
\def\pmbb#1{\setbox0\hbox{${#1}$} \copy0 \kern-\wd0
      \kern.2pt \copy0 \kern-\wd0 \kern.2pt \box0}
\def\pmbbb#1{\setbox0\hbox{${#1}$} \copy0 \kern-\wd0
      \kern.2pt \copy0 \kern-\wd0 \kern.2pt
    \copy0 \kern-\wd0 \kern.2pt \box0}
\def\pmxb#1{\setbox0\hbox{${#1}$} \copy0 \kern-\wd0
      \kern.2pt \copy0 \kern-\wd0 \kern.2pt
      \copy0 \kern-\wd0 \kern.2pt \copy0 \kern-\wd0 \kern.2pt \box0}
\def\pmxbb#1{\setbox0\hbox{${#1}$} \copy0 \kern-\wd0 \kern.2pt
      \copy0 \kern-\wd0 \kern.2pt
      \copy0 \kern-\wd0 \kern.2pt \copy0 \kern-\wd0 \kern.2pt
      \copy0 \kern-\wd0 \kern.2pt \box0}

\def\sJ{{\sss J}}

\def\sP{{\sss P}}
\def\sZ{{\sss Z}}

\def\sT{{\sss T}}
\def\sV{{\sss V}}

\def\sv{{\sss v}}

\def\sj{{\sss j}}

\mathchardef\za="710B  
\mathchardef\zb="710C  
\mathchardef\zg="710D  
\mathchardef\zd="710E  
\mathchardef\zve="710F 
\mathchardef\zz="7110  
\mathchardef\zh="7111  
\mathchardef\zvy="7112 
\mathchardef\zi="7113  
\mathchardef\zk="7114  
\mathchardef\zl="7115  
\mathchardef\zm="7116  
\mathchardef\zn="7117  
\mathchardef\zx="7118  
\mathchardef\zp="7119  
\mathchardef\zr="711A  
\mathchardef\zs="711B  
\mathchardef\zt="711C  
\mathchardef\zu="711D  
\mathchardef\zvf="711E 
\mathchardef\zq="711F  
\mathchardef\zc="7120  
\mathchardef\zw="7121  
\mathchardef\ze="7122  
\mathchardef\zy="7123  
\mathchardef\zf="7124  
\mathchardef\zvr="7125 
\mathchardef\zvs="7126 
\mathchardef\zf="7127  
\mathchardef\zG="7000  
\mathchardef\zD="7001  
\mathchardef\zY="7002  
\mathchardef\zL="7003  
\mathchardef\zX="7004  
\mathchardef\zP="7005  
\mathchardef\zS="7006  
\mathchardef\zU="7007  
\mathchardef\zF="7008  
\mathchardef\zW="700A  

\def\cP{\mathcal{P}}

\def\cM{\mathcal{M}}

\def\bl({\left(}
\def\br){\right)}
\def\R{\mathbb R}

\def\bt{\begin{tiny}}
\def\et{\end{tiny}}

\def\ba{\begin{array}}
\def\ea{\end{array}}

\def\ti{\times}
\def\xd{\mathrm{d}}
\def\uxd{{\underline{\mathrm{d}}}}
\def\xdv{\mathrm{d}^\sv}

\def\pa{\partial}
\def\zx{\xi}
\def\ot{\otimes}
\def\Aff{\sss{Aff}}

\def\Hom{\sss{Hom}}

\def\Sec{\sss{Sec}}

\newcommand{\beas}{\begin{eqnarray*}}
\newcommand{\eeas}{\end{eqnarray*}}

\newdir{|>}{%
!/4.5pt/@{|}*:(1,-.2)@^{>}*:(1,+.2)@_{>}}
\newdir{ (}{{}*!/-5pt/@^{(}}
\newtheorem{thm}{Theorem}[section]
\newtheorem{theorem}{Theorem}[section]

\theoremstyle{definition}

\newtheorem{remark}[thm]{Remark}
\newtheorem{definition}[thm]{Definition}

\begin{document}
\pagestyle{myheadings}
\markboth{Katarzyna Grabowska and Janusz Grabowski}{From statics to field theory}

\title{Tulczyjew Triples: From Statics to Field Theory\thanks{Research  founded by the  Polish National Science Centre grant
under the contract number DEC-2012/06/A/ST1/00256.}
}
\date{}

\maketitle

\vskip-1cm
\centerline{\scshape Katarzyna Grabowska}
\medskip
{\footnotesize
 \centerline{Physics Department,
                University of Warsaw}
   \centerline{Ho\.za 69, 00-681 Warszawa, Poland}
   \centerline{E-mail: konieczn@fuw.edu.pl}
\medskip

\centerline{\scshape Janusz Grabowski}
\medskip
{\footnotesize
 \centerline{Institute of Mathematics, Polish Academy of Sciences}
   \centerline{\'Sniadeckich 8, 00-956 Warszawa, Poland}
   \centerline{E-mail: jagrab@impan.pl}
}

\begin{abstract}
We propose a geometric approach to dynamical equations of physics, based on the idea of the Tulczyjew triple. We show the evolution of these concepts, starting with the roots lying in the variational calculus for statics, through Lagrangian and Hamiltonian mechanics, and concluding with Tulczyjew triples for classical field theories, illustrated with a numer of important examples.
\end{abstract}

\section{Introduction}\label{intro}

Variational calculus is a natural language for describing statics of mechanical systems. All mathematical
objects that are used in statics have direct physical interpretations. Similar mathematical tools
are widely used also in other theories, like dynamics of particles or field theories, however the
links between mathematical language and physical system are in these cases more sophisticated.

In classical mechanics, variational calculus was used first for deriving equations of motion of mechanical systems in the configuration space, i.e. the Euler-Lagrange equations. In most frameworks (e.g. the Klein's approach), deriving the Euler-Lagrange equations is the main objective.
On the other hand, in numerous works by W. M. Tulczyjew (for example in the book \cite{Tu3} and papers \cite{Tu6,Tu7,Tu9,Tu10}) one may find another philosophy of using variational calculus in mechanics and field theories in which the phase dynamics plays a fundamental role. This philosophy, leading to the geometrical structure known as the {\it Tulczyjew triple}, is being more and more recognized by many theoretical physicists and mathematicians.

The Tulczyjew triple has proved to be very useful in describing mechanical systems, including those with singular Lagrangians or subject to constraints \cite{TU1}. Starting from basic concepts of variational calculus, we will construct the Tulczyjew triple for first-order Field Theory. The important feature of our approach is that we do not postulate {\it ad hoc} the ingredients of the theory, but obtain them as unavoidable consequences of the variational calculus. This picture of Field Theory is covariant and complete, containing not only the Lagrangian formalism and Euler-Lagrange equations but also the phase space, the phase dynamics and the Hamiltonian formalism. Since the configuration space turns out to be an affine bundle, we have to use
affine geometry, in particular the notion of affine duality and affine phase space. In our formulation,
the two maps $\alpha$ and $\beta$ which constitute the Tulczyjew triple are morphisms of double structures of affine-vector bundles. We discuss also the Legendre transformation, i.e. the transition between the Lagrangian and the Hamiltonian formulation of the first-order field theory.

In this survey, based on \cite{G}, we will present the Tulczyjew triple for first-order field theories in a very general setting, i.e. in the case where fields are sections of some differential fibration, with no additional structure assumed. Our paper is organized as follows.

Since we have to use some affine geometry, we start, in section \ref{sec:1}, with a short sketch of some affine constructions and theorem \ref{th:1} describing a canonical isomorphism of  certain phase spaces. Then, we present variational calculus in statics (section \ref{sec:3}) and its application to mechanics (sections \ref{sec:3a} and \ref{sec:4}). In section \ref{sec:7}, we describe the classical Tulczyjew triple for mechanics, and in section \ref{sec:7a} for mechanics on algebroids. In the next step, we pass to field theory. Like in mechanics, we have to start from some field theoretical construction for a bounded domain to find correct mathematical representations for certain physical quantities (section \ref{sec:9}). Lagrangian and Hamiltonian sides of the field-theoretical triple are constructed in sections \ref{sec:11} and \ref{sec:12} respectively. The remaining sections are devoted to examples. There is also an appendix containing the proof of theorem \ref{th:1}.

Note finally that classical field theory is usually associated with the concept of a multisymplectic structure.  The multisymplectic approach appeared first in the papers of the `Polish school' \cite{Ga,KS,KT,Tu5}. Then, it was developed by Gotay, Isennberg, Marsden, and others in \cite{GIMa,GIMb}. The original idea of the multisymplectic structure has been thoroughly investigated and developed by many contemporary authors, see e.g. \cite{CIL,CCI1,EM,FP1,FP2}. The Tulczyjew triple in the context of multisymplectic field theories appeared recently in \cite{LMS} and \cite{CGM}. A similar picture, however with differences on the Hamiltonian side, one can find in \cite{GM} (see also \cite{GMS,Kr}).

\section{Affine phase spaces}\label{sec:1}
Affine geometry turned out to be an important tool in mechanics and field theory. Let us begin with a short review of affine structures that will be needed later on. Details and further fundamental observations can be found e.g. in \cite{GGU1}.

Let $A$ be an affine space modeled on a vector space $\sv(A)$.
This means that the commutative group $\sv(A)$ acts freely and transitively on $A$ by addition
$$A\ti\sv(A)\ni (a,v)\mapsto a+v\,.$$
In other words, the naturally defined differences $a_1-a_2$ of points of $A$ belong to $\sv(A)$.
On affine spaces there are defined \emph{affine combinations} of points, $ta_1+(1-t)a_2$, for all $a_1,a_2\in A$ and $t\in\R$. Note that {\it convex combinations} are those affine combinations
$ta_1+(1-t)a_2$ for which $0\le t\le 1$.

All this can be extended to affine bundles $\zt:A\to N$ modelled on a vector bundle $\sv(\zt):\sv(A)\to N$. Any vector bundle is an affine bundle and fixing a section $a_0$ of $A$ induces an isomorphism of affine bundles $A$ and $\sv(A)$,
$$\sv(A)\ni v\mapsto a_0+v\in A\,.$$
Using coordinates $(x^i)$ in the open set $\mathcal{O}\subset N$, a local section $a_0: \mathcal{O}\rightarrow A$,
and local base of sections $e_a:\mathcal{O}\rightarrow \sv(A)$, we can construct an adapted coordinate
system $(x^i, y^a)$ in $\tau^{-1}(\mathcal{O})$. An element $a\in A$ can be written as
$a=a_0(\tau(a))+y^ae_a(\tau(a))$.
\begin{definition}\label{def:av}
An \emph{AV-bundle} is an affine bundle $\zz:\sZ\to \cM$ modeled on a trivial one-dimensional vector bundle $\cM\ti Y$, where $Y$ is a one-dimensional vector space. In applications $Y$ will be either $\R$ or the one dimensional vector space of top forms on a manifold.
\end{definition}
For the affine space $A_q=\tau^{-1}(q)$, we consider its {\it affine dual}, i.e.
the space $A_q^\dag(Y)$ of all affine maps from $A_q$ to the one-dimensional vector space $Y$.
\begin{definition}\label{def:affdual}
The bundle $\zt^\dag:A^\dag(Y) \longrightarrow N$, where $A^\dag(Y)=\Aff(A,Y)$ is
the set of all affine maps on fibres of $\zt$, is called the \emph{affine dual bundle} with values in $Y$. Instead of $A^\dag(\R)$ we will write simply $A^\dag$.
\end{definition}
Every affine map $\phi:A_1\to A_2$ has a well-defined {\it linear part},  $\sv(\phi):\sv(A_1)\to\sv(A_2)$, therefore there is a projection
\begin{equation}\label{theta}\zvy: A^\dag(Y)\longrightarrow \sv(A)^\ast\otimes_N (N\times Y)=\Hom(\sv(A),Y).\end{equation}
The above bundle is a canonical example of an AV-bundle which is modeled on
\begin{equation}\label{hb}
(\sv(A)^\ast\otimes_N (N\times Y))\ti_NY\,.
\end{equation}
In the following we shall write $\sv(A)^\ast\otimes_N Y$ instead of $\sv(A)^\ast\otimes_N (N\times Y)$
to simplify the notation. The fibre of $\sv(A)^\ast\otimes_N Y$ over a point $x\in N$ is $\sv(A_x)^\ast\otimes Y$.

Using the dual base sections
$\ze^a:\mathcal{O}\rightarrow \sv(A)^\ast$ and a base element $u$ of $Y$, we construct
an adapted coordinate system $(x^i, p_a, r)$ on $(\zt^\dag)^{-1}(\mathcal{O})$.
An affine map $\varphi$ on $A_q$ can be written as $\varphi(a)=(p_a\ze^a(a-a_0(q))+r)u$.
The map $\zvy$ in coordinates reads $(x^i,p_a,r)\mapsto(x^i,p_a)$.

In many constructions functions on a manifold can be replaced by sections of an AV-bundle
over that manifold. We can obtain also an affine analog of the differential of a function
and an affine version of the cotangent bundle as follows. Given an AV-bundle $\zz:\sZ\to \cM$ and
$F_1,F_2\in\Sec(\sZ)$, $F_1-F_2$ may be seen as a map
$$F_1-F_2:\cM\to Y\,,$$
so the differential
$$\xd(F_1-F_2)(m)\in Y$$
is well defined.
\begin{definition}\label{def:phase}
The \emph{phase bundle} $\sP\sZ$ of an AV-bundle $\sZ$ is the affine bundle of cosets $\uxd F(m)=[(m,F)]$
(`affine differentials') of the equivalence relation
\begin{equation*}(m_1,F_1)\sim(m_2,F_2)\ \Leftrightarrow\ m_1=m_2\,,\quad \xd(F_1-F_2)(m_1)=0\,.\end{equation*}
\end{definition}

Fixing a section $F_0:\cM\to \sZ$ and a basic vector $u^\ast\in Y^\ast$, we get a diffeomorphism
\begin{equation*}\psi:\sP\sZ\to\sT^*\cM\,, \quad \uxd F(m)\mapsto \xd(u^\ast(F-F_0))(m)\,.\end{equation*}
As the canonical symplectic form on $\sT^*\cM$ is linear and invariant with respect to translations by closed 1-forms, its pull-back does not depend on the choice of $F_0$ nor $u^\ast$,
and turns $\sP\sZ$ into a canonically symplectic manifold.

Now, let us consider a finite-dimensional vector bundle $V$ over a manifold $N$ and choose a vector subbundle $W$ over $N$. The bundle $\tau: V\rightarrow V\slash W$,
where $\tau$ is the canonical projection from $V$ onto the quotient bundle $V\slash W$, is an affine bundle modeled
on the trivial bundle
$$\sv(\tau)=pr_1:V\slash W\times_N W\rightarrow V\slash W\,.$$ We can consider therefore its affine
dual $V^\dag_W\rightarrow V\slash W$. We observe that the bundle $$V^\dag_W\rightarrow V\slash W\times_N W^\ast$$ is an AV-bundle. We know that the corresponding phase bundle is $\sP V^\dag_W$ is canonically a symplectic manifold which, somehow unexpectedly, can be identified as follows.

\begin{theorem}\label{th:1} There is a canonical symplectomorphism $\sP V^\dag_W\simeq \sT^*V$. In particular, if \ $W\subset V$ are just vector spaces, i.e. vector bundles over single points, we have $\sP V^\dag_W\simeq V\times V^\ast$.
\end{theorem}

The proof of the theorem can be found in the Appendix.

\section{Variational calculus}
\label{sec:2}

\subsection{Statics}
\label{sec:3}
Variational calculus used in mechanics and field theory is based on ideas from statics.
We assume that the set of configurations of the static system we describe is a differential manifold
$Q$.
\begin{pict}[ht]
\begin{floatrow}[2]
\floatbox{pict}[0.47\textwidth]
{\caption{Static system...}\label{fig:1}}
{\centering\includegraphics[width=0.46\textwidth]{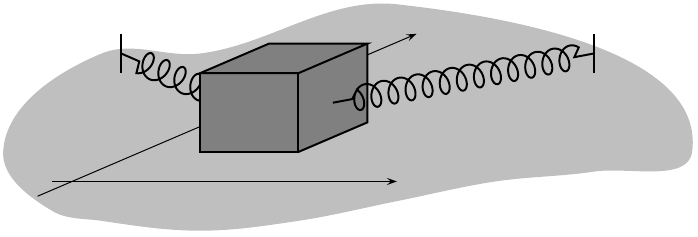}}
\floatbox{pict}[0.47\textwidth]
{\caption{and its mathematical model}\label{fig:2}}
{\centering\includegraphics[width=0.35\textwidth]{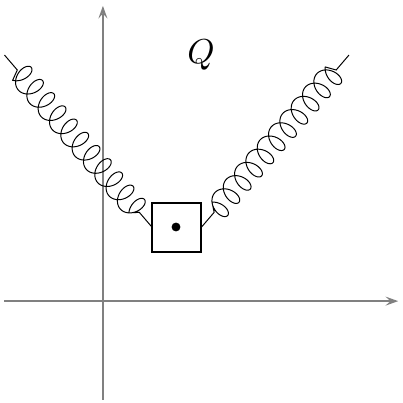}}
\end{floatrow}
\end{pict}
We are usually interested in equilibrium configurations of an isolated system,
as well as a system with an interaction with other static systems. The system alone or in interaction is
examined by preforming processes and calculating the cost of each process. We assume that all the processes
are {\it quasi-static}, i.e. they are slow enough to produce negligible dynamical effects. Every process can be
represented by a one-dimensional smooth oriented submanifold with boundary (Fig.\ref{fig:3}).
\begin{pict}[ht]
\begin{floatrow}[1]
\floatbox{pict}[0.47\textwidth]
{\caption{Quasistatic processes}\label{fig:3}}
{\centering\includegraphics[width=.4\textwidth]{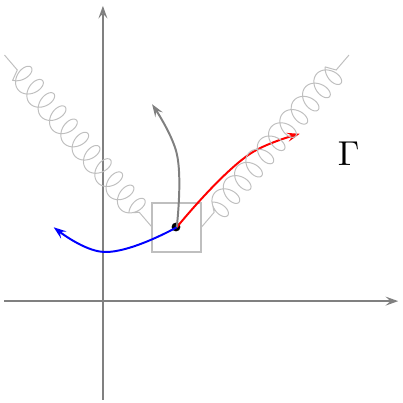}}
\end{floatrow}
\end{pict}
It may happen that, for some
reasons, not all the processes are admissible, i.e. the system is constrained. All the information about the
system is therefore encoded in the three objects: the configuration manifold $Q$, the set of all admissible
processes, and the cost function that assigns a real number to every process. The cost function should fulfill
some additional conditions, e.g. it should be additive in the sense that if we break a process into two
subprocesses, then the cost of the whole process should be equal to the sum of the costs of the two
subprocesses. Usually we assume that the cost function is local, i.e. for each process it is an integral of a
certain positively homogeneous function $W$ on $\sT Q$ or, in case of constrained system, on some
subset $\Delta\subset\sT Q$. Vectors tangent to admissible processes
are called {\it admissible virtual displacements} (Fig.\ref{fig:4}).
\begin{pict}[ht]
\begin{floatrow}[1]
\floatbox{pict}[0.47\textwidth]
{\caption{Virtual displacements}\label{fig:4}}
{\centering\includegraphics[width=.4\textwidth]{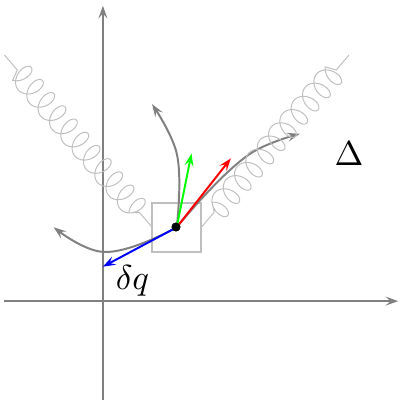}}
\end{floatrow}
\end{pict}
\begin{definition}\label{def:equilibrium}
Point $q\in Q$ is an {\it equilibrium point} of the system if for all processes starting in $q$ the cost function
is non-negative, at least initially.
\end{definition}
The first-order necessary condition for the equilibrium says that a
point $q$ is an equilibrium point of the system if
$$W(\delta q)\geq 0$$
for all vectors $\delta q\in \Delta$.

Interactions between systems are described by composite systems. We can compose systems that have the
same configuration space $Q$. The composite system (Fig.\ref{fig:5}) is described by the intersection of the sets of admissible
processes and the sum of the cost functions $W=W_1+W_2$. From now on, the subscript $1$ will denote `our system'
and the subscript $2$ an external system we use for collecting information about our system.
The interaction with an external system is usually described in terms
of forces $\varphi\in\sT^*Q$.
\begin{pict}[ht]
\begin{floatrow}[1]
\floatbox{pict}[0.47\textwidth]
{\caption{Composite system}\label{fig:5}}
{\centering\includegraphics[width=.4\textwidth]{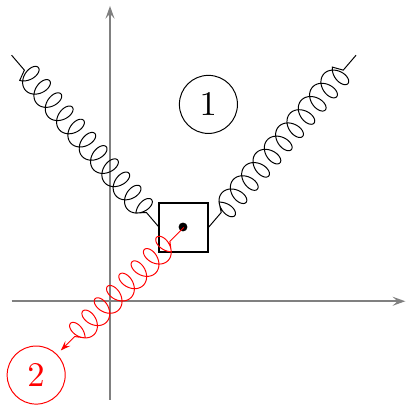}}
\end{floatrow}
\end{pict}
The forces can be understood as distinguished systems, called {\it regular},
for which all the processes are admissible and the function $W_2$ is the differential of a certain function
$U:Q\rightarrow \R$. A regular system at a point $q$ is represented by $\varphi=-\xd U(q)$. The `minus' sign comes
from the fact that the system is external and changing its configuration has just the opposite cost for us.
\begin{definition}\label{def:constitutive}
The subset $\mathcal{C}\subset\sT^\ast Q$ of all external forces which are in equilibrium with our system is called \emph{the constitutive set}.
\end{definition}
If our system is regular, i.e. $W_1(\zd q)=\langle\xd U,\zd q\rangle$ for a function $U:Q\to\R$, then the constitutive set is $\mathcal{C}=\xd U(Q)$.
In mechanics and field theory the most important type of systems are analogs of regular systems and regular systems with constraints.
To apply the ideas coming from statics to other theories, we shall specify for every theory
\begin{itemize}
\item configurations $Q$,
\item processes (or at least infinitesimal processes),
\item functions on $Q$  (to define regular systems),
\item covectors $\sT^\ast Q$ (to define constitutive sets).
\end{itemize}

\subsection{Mechanics for finite time interval}
\label{sec:3a}

Let $M$ be a manifold of positions of a mechanical system. We will use smooth paths in $M$
and first-order Lagrangians $L:\sT M\to\R$. We also fix the time interval $[t_0,t_1]$.
{\it Configurations} $q$ are pieces of smooth paths (Fig.\ref{fig:6})
$$q:[t_0, t_1]\rightarrow M.$$
\begin{pict}[ht]
\begin{floatrow}[1]
\floatbox{pict}[0.47\textwidth]
{\caption{A configuration}\label{fig:6}}
{\centering\includegraphics[width=.35\textwidth]{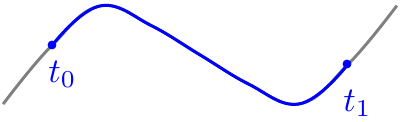}}
\end{floatrow}
\end{pict}\noindent The set of all configurations will be denoted by $Q$. Since $Q$ is not a standard manifold, we have to introduce `by hand' the concepts of process in $Q$, a smooth function on $Q$, and vector tangent to $Q$. Smooth functions will be associated with Lagrangians, i.e. for any
function $L:\sT M\to\R$ we define an action functional $S:Q\rightarrow \R$ by
$$S(q)=\int_{t_0}^{t_1} L(\dot q)dt.$$
Parameterized processes (Fig.\ref{fig:7}) in $Q$ come from homotopies $q_s(t)=\chi(s,t)$, i.e. smooth maps
$$\chi:\R^2\supset I\times J\rightarrow M\,,$$
where $I$ is some neighborhood of zero in $\R$ and $J$ contains $[t_0, t_1]$.
\begin{pict}[ht]
\begin{floatrow}[1]
\floatbox{pict}[0.4\textwidth]
{\caption{A curve in configurations}\label{fig:7}}
{\centering\includegraphics[width=.2\textwidth]{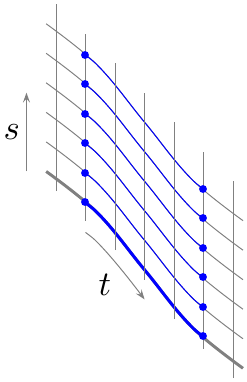}}
\end{floatrow}
\end{pict}
Smooth curves and smooth functions are defined in such a way that the composition of a curve
with a function is a real smooth function, smooth in usual sense. We can therefore employ the standard definition
of tangent vectors and covectors as equivalence classes of curves and equivalence classes of functions
respectively.
\begin{definition}\label{def:vectors}
A {\it vector tangent to $Q$} is an equivalence class of smooth curves with respect to the equivalence relation which
says that two curves $q_s$ and $q'_s$ are equivalent if, for $t\in[t_0,t_1]$, $q_0(t)=q'_0(t)$ and,
for all smooth functions $S$,
$$\frac{\xd}{\xd s}_{|s=0}S(q_s)=\frac{\xd}{\xd s}_{|s=0}S(q'_s).$$
\end{definition}
\begin{definition}\label{def:functions}
A {\it covector tangent to $Q$} is an equivalence class of pairs $(q,S)$ with respect to the equivalence relation which
says that two pairs $(q,S)$ and $(q',S')$ are equivalent if, for $t\in[t_0,t_1]$, $q(t)=q'(t)$ and,
for all smooth curves $s\mapsto q_s$ such that $q_0=q$, we have
$$\frac{\xd}{\xd s}_{|s=0}S(q_s)=\frac{\xd}{\xd s}_{|s=0}S'(q_s).$$
\end{definition}
Since tangent vectors and covectors defined as equivalence classes are abstract objects hard to work with,
we need some convenient representations for them. Performing integration by parts, as when deriving Euler Lagrange equations, we get
\begin{equation}\label{eq:convenient}
\left.\frac{d}{ds}\right|_{s=0}S(q_s)=
\int_{t_0}^{t_1}\langle\mathcal{E}L(\ddot q), \delta q\rangle dt+ \left.\frac{}{}\langle\,\cP L(\dot q),\delta q\,\rangle\right|^{t_1}_{t_0}\,,
\end{equation}
where
$\mathcal{E}L:\sT^2M\to\sT^*M$ and $\cP L=\xdv L:\sT M\to\sT^*M$ are bundle maps and
$\delta q:[t_0, t_1]\rightarrow \sT M$ is a curve in $\sT M$ whose value at $t$ is a vector tangent to the
curve $s\mapsto q_s(t)$ (Fig.\ref{fig:8}). It is easy to see that tangent vectors are in a one-to-one correspondence
with paths $\zd q$ in $\sT M$, and covectors are in a one-to-one correspondence with triples $(f,p_0,p_1)$,
$f:[t_0,t_1]\rightarrow\sT^\ast M$,
$p_i\in\sT^\ast_{q(t_i)}M$ (Fig.\ref{fig:9}).
\begin{pict}[ht]
\begin{floatrow}[2]
\floatbox{pict}[0.47\textwidth]
{\caption{Tangent vectors}\label{fig:8}}
{\centering\includegraphics[width=0.2\textwidth]{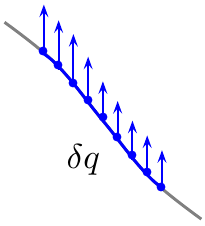}}
\floatbox{pict}[0.47\textwidth]
{\caption{Covectors}\label{fig:9}}
{\centering\includegraphics[width=0.2\textwidth]{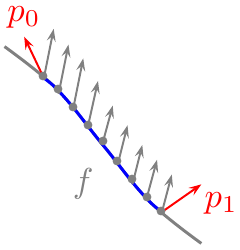}}
\end{floatrow}
\end{pict}
We have found another representation of covectors which is referred to by Tulczyjew and coworkers \cite{TU2} as a {\it Liouville structure},
$$\alpha: \mathbb{P}Q=\{(f,p_0,p_1)\}\longrightarrow \sT^\ast Q\,. $$
The mechanical system with Lagrangian $L$ is, from a statical point of view, a regular system with
cost function given by $\xd S$. The constitutive set is therefore $\mathcal{C}=\xd S(Q)$. We prefer,
however, to use convenient representations and to call the appropriate set the dynamics of a system.
\begin{definition}\label{def:dynamic}
The {\it (phase) dynamics} of a mechanical system is a subset $\mathcal{D}$ of  $\mathbb{P}Q=\{(f,p_0,p_1)\}$ defined by
$$\mathcal{D}=\alpha^{-1}(dS(Q)),$$
i.e.,
$$\mathcal{D}=\left\{(f,p_0,p_1):\;\;
f(t)=\mathcal{E}L(\ddot q(t)),\quad
p_a=\cP L(\dot q(t_a))\,,\ a=0,1\right\}\,.$$
\end{definition}

\noindent Explicitly, writing in coordinates, $q=(x^i(t))$, $\dot q=(x^i(t),\dot x^j(t))$, we have
$$f_i(t)=\frac{\partial L}{\partial x^i}(\dot q(t))-\frac{\xd}{\xd t}\left(\frac{\partial L}{\partial \dot{x}^i}(\dot q(t))\right)\,,\quad (p_a)_i=\frac{\partial L}{\partial \dot{x}^i}(\dot q(t_a))\,,\ a=0,1\,.$$

\subsection{Mechanics: Infinitesimal version}
\label{sec:4}

Mechanics for finite time interval is very useful for creating intuitions but not particularly convenient for analyzing the behavior of the system. For the latter purpose, we need actual differential equations for curves in the phase space and in the configuration space. We can get them by passing to the infinitesimal
formulation of mechanics in which $M$ will stand as a manifold of positions of mechanical system and Lagrangians will be of the first order. As configurations we choose now `infinitesimal pieces of paths', i.e.
$Q=\sT M$, if $x:\R\rightarrow M$ is a smooth curve, then $q=\dot x(0)$.
\begin{pict}[ht]
\begin{floatrow}[1]
\floatbox{pict}[0.4\textwidth]
{\caption{Infinitesimal configuration}}
{\centering\includegraphics[width=.3\textwidth]{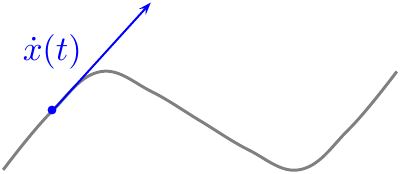}}
\end{floatrow}
\end{pict}
This time, the configuration space is a finite-dimensional manifold, therefore we know what are smooth curves
and functions, as well as tangent and cotangent spaces. If the system is regular, its cost function
is given by $\xd L$ and its constitutive set is just $\mathcal{C}=\xd L(\sT M)\subset\sT^\ast\sT M$.
However, it is interesting what we get out of convenient representatives of vectors and covectors,
while passing to the infinitesimal formulation. Let us start with tangent vectors. Vectors tangent
to the space of infinitesimal configurations are elements $\delta q$ of $\sT\sT M$, i.e. vectors
tangent to curves in $\sT M$. Starting from a homotopy $(t,s)\mapsto \chi(t,s)$, we get
a configuration $q=\dot\chi(0,0)$ which is the vector tangent to the curve $t\mapsto \chi(t, 0)$
at $t=0$, and the curve $s\mapsto \dot\chi(0,s)$, where $\dot\chi(0,s)$ is the vector tangent to
the curve $t\mapsto \chi(t, s)$ at $t=0$ (Fig.\ref{fig:12}).
\begin{pict}[ht]
\begin{floatrow}[1]
\floatbox{pict}[0.5\textwidth]
{\caption{A curve in infinitesimal configuration}}
{\centering\includegraphics[width=.2\textwidth]{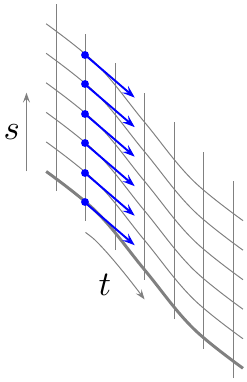}}
\end{floatrow}
\end{pict}
But in (\ref{eq:convenient})
we did the other way around, i.e. we first differentiated with respect to $s$, obtaining
the curve $t\mapsto \delta\chi(t, 0)$ with values in $\sT M$ (Fig.\ref{fig:13}), and then with respect to $t$.
What we have just described is the well-known canonical involution
\begin{equation}\label{eq:kappa}
\kappa_M:\sT\sT M\longrightarrow\sT\sT M,\qquad \delta\dot\chi(0,0)\longmapsto\left(\delta\chi\right)^\cdot(0,0)\,.
\end{equation}
A convenient representation for a tangent vector $\delta q=\delta\dot\chi(0,0)$ (infinitesimal variation) is another element
of $\sT\sT M$, namely $\kappa_M(\delta q)=\left(\delta\chi\right)^\cdot(0,0)$.
\begin{pict}[ht]
\begin{floatrow}[2]
\floatbox{pict}[0.47\textwidth]
{\caption{A curve in configuration and...}\label{fig:12}}
{\centering\includegraphics[width=0.2\textwidth]{rys11.pdf}}
\floatbox{pict}[0.47\textwidth]
{\caption{...a curve in variation.}\label{fig:13}}
{\centering\includegraphics[width=0.2\textwidth]{rys8.pdf}}
\end{floatrow}
\end{pict}

Covectors are of course elements of $\sT^\ast\sT M$ and the constitutive set $\mathcal{C}$ is a subset of $\sT^\ast\sT M$ given by the differential of a Lagrangian, provided the system is not constrained. However, we can find a better description of $\mathcal{C}$ in terms of convenient representations of covectors.
For infinitesimal time interval, the formula (\ref{eq:convenient}) reads
\begin{equation}\label{eq:infinitesimal}
\langle\xd L, \delta q\rangle=\langle \mathcal{E}L(\ddot\chi(0,0)), \delta \chi(0,0)\rangle+
\frac{\xd}{\xd t}_{|t=0}\langle\mathcal{P}L(\dot\chi(t,0)),\delta \chi(t,0)\rangle\,.
\end{equation}
On the left hand side there is an evaluation of the covector $\xd L$ on the variation $\delta q$
of the infinitesimal configuration $q$, while on the right hand side we have the external force
$f(0)=\mathcal{E}L(\ddot\chi(0,0))$ evaluated on the variation of the position $\delta \chi(0,0)$
and the second term involving the momentum. Let us assume that there are no external forces.
The curve $p: t\mapsto \mathcal{P}L(\dot\chi(t,0))$ gives values in $\sT^\ast M$, while
the curve $\gamma: t\mapsto \delta \chi(t,0)$ gives values in $\sT M$. The value of
$\frac{\xd}{\xd t}_{|t=0}\langle\mathcal{P}L(\dot\chi(t,0)),\delta \chi(t,0)\rangle$
depends on values of the curves $\gamma$ and $p$ at $0$ and on vectors tangent to those curves.
It can be understood as a coupling between two vector bundles
$\sT\tau_M: \sT\sT M\rightarrow \sT M$ and $\sT\pi_M: \sT\sT^\ast M\rightarrow \sT M$,
\begin{equation}\label{eq:eval}
\langle\!\langle \dot p, (\delta \chi)^{\cdot}\rangle\!\rangle=\left.\frac{d}{dt}\right|_{t=0}\langle p(t), \delta\chi(t,0)\rangle.
\end{equation}
The bundle $\sT\pi_M$ is now dual to $\sT\tau_M$. Since $\kappa_M$ is an isomorphism between
$\sT\tau_M$ and $\tau_{\sT M}$, we can find the dual isomorphism between appropriate dual bundles, namely
\begin{equation}\label{eq:alpha}
\alpha_M:\sT\sT^\ast M \longrightarrow \sT^\ast\sT M\,.
\end{equation}
Formula (\ref{eq:infinitesimal}) says that a covector $\xd L(\dot\gamma)$ can be conveniently represented
by a pair $(f,\dot p)$. If external force is equal to zero, then $\dot p$ and $\xd L(\dot\gamma)$ are related by
$\alpha_M$. The infinitesimal dynamics of a system with no external forces is therefore
$$\sT\sT^\ast M\supset \mathcal{D}=\alpha_M^{-1}(\xd L(\sT M)).$$
Since $\mathcal{D}$ is a subset of the tangent space, it can be regarded as an (implicit) first-order
differential equation for curves in the phase space $\sT M$.

\section{Tulczyjew triples}
\label{sec:5}

In sections \ref{sec:3a} and \ref{sec:4} we have discussed the results of using statical ideas in mechanics. The mechanics for
finite time interval provided us with concepts of convenient representations of vectors and covectors. For the
infinitesimal time interval, we have obtained a way of generating differential equations describing dynamics of
a system in the phase space. The results of section \ref{sec:4} can be formulated in an elegant way as the Lagrangian
side of the Tulczyjew triple. The Tulczyjew triple for mechanics is built out of maps between $\sT\sT^\ast M$,
$\sT^\ast\sT^\ast M$, and $\sT^\ast\sT M$ which are examples of double vector bundles.

\subsection{Double vector bundles}
\label{sec:6}

The following geometric definition (cf. \cite{GR,GR1}) is a simplification of
the original categorical concept of a double vector bundle due to
Pradines \cite{Pr1}, see also \cite{KU,Ma}.

\begin{definition} A \emph{double vector bundle} is a manifold with two compatible vector bundle structures.
Compatibility means that the Euler vector fields (generators of homothethies), associated with the two structures, commute.
\end{definition}
This definition implies that, with every double vector bundle, we
can associate the following diagram of vector bundles in which
both pairs of parallel arrows form vector bundle morphisms:
\begin{equation}\label{F1.3x}\xymatrix{
 & K\ar[dr]^{\tau_2}\ar[dl]_{\tau_1} & \\
K_1\ar[dr]^{\tau'_2} &  & K_2\ar[dl]_{\tau'_1} \\
 & M & }
\end{equation}

The first example of double vector bundle mentioned in these notes is $\sT\sT M$ with two projections over $\sT M$: the
canonical one, $\zt_{\sT M}$, which associates to a vector tangent to $\sT M$ the point in $\sT M$ where the vector is attached,
and the tangent one, $\sT\tau_M$, which associates to a vector tangent to $\sT M$ its tangent projection on $\sT M$.  Local
coordinates $(x^i)$ in an open subset $\mathcal{O}\subset M$ can be used to define adapted coordinates $(x^i,\dot x^j)$
in $\tau_M^{-1}(\mathcal{O})$ and $(x^i,\dot x^j,\zd x^k,\delta\dot{x}^l)$ in an appropriate subset of $\sT\sT M$.
The vector fields $\nabla_1$ and $\nabla_2$ are the two commuting Euler vector fields for the two compatible vector
bundle structures in $\sT\sT M$.

\begin{minipage}[c]{0.45\linewidth}
$$\xymatrix@R+5pt{
 & \sT\sT M\ar[dr]^{\sT\zt_M}\ar[dl]_{\zt_{\sT M}} & \\
\sT M\ar[dr]^{\zt_M} & & \sT M\ar[dl]^{\zt_M} \\
 & M &
}$$
\end{minipage}
\hspace{0.5cm}
\begin{minipage}[c]{0.45\linewidth}
\begin{align*}
\nabla_1&=\zd{x}^i\partial_{\zd{x}^i}+\delta\dot{x}^j\partial_{\delta\dot{x}^j}\,, \\
\nabla_2&=\dot{x}^i\partial_{\dot{x}^i}+\delta\dot{x}^j\partial_{\delta\dot{x}^j}\,.
\end{align*}
\end{minipage}

The diffeomorphism $\zk_M:\sT\sT M\to\sT\sT M$ interchanges the two vector bundle structures,
$$(x^i,\dot{x}^j,\zd x^k,\delta\dot{x}^l)\mapsto(x^i,\zd x^k,\dot{x}^j,\delta\dot{x}^l)\,.$$
It contains also the information about the bracket of vector fields.

More general examples of double vector bundles are: the tangent bundle $\sT E$ and the cotangent bundle $\sT^\ast E$
for a vector bundle $\zt: E\rightarrow M$. In coordinates $(x^i, y^a)$ in $E$
and $(x^i, y^a, \dot x^j, \dot y^b)$ in $\sT E$, we get

\begin{minipage}[c]{0.45\linewidth}
$$\xymatrix@R+5pt{
 & \sT E\ar[dr]^{\sT\zt}\ar[dl]_{\tau_{E}} & \\
E\ar[dr]^{\zt} & & \sT M\ar[dl]^{\tau_M} \\
 & M &
}$$
\end{minipage}
\hspace{0.5cm}
\begin{minipage}[c]{0.45\linewidth}
\begin{align*}
\tau_{E}:\sT E&\longrightarrow E \\
(x^i,y^a, \dot x^j,\dot y^b)&\longmapsto (x^i,y^a) \\
\quad\\
\sT\tau: \sT E&\longrightarrow \sT M \\
(x^i,y^a, \dot x^j,\dot y^b)&\longmapsto (x^i,\dot x^j)
\end{align*}
\end{minipage}

\medskip\noindent For the cotangent bundle $\sT^\ast E$, we use coordinates $(x^i,y^a,p_j,\xi_b)$.
The structure of $\sT^\ast E$ as a double vector bundle is the following:

\begin{minipage}[c]{0.45\linewidth}
$$\xymatrix@R+5pt{
 & \sT^\ast E\ar[dr]^{\zz_E}\ar[dl]_{\zt_{E}} & \\
E\ar[dr]^{\zt} & & E^*\ar[dl]^{\pi} \\
 & M &
}$$
\end{minipage}
\hspace{0.5cm}
\begin{minipage}[c]{0.45\linewidth}
\begin{align*}
\pi_E: \sT^\ast E&\longrightarrow E \\
(x^i,y^a,p_j,\xi_b)&\longmapsto (x^i,y^a)\\
\quad \\
\zz_E: \sT^\ast E&\longrightarrow E^* \\
(x^i,y^a,p_j,\xi_b)&\longmapsto (x^i,\xi_a)
\end{align*}
\end{minipage}

\medskip\noindent The projection $\zz_E$ is constructed as follows. Let us observe that vectors tangent to
the fibre of a vector bundle can be identified with elements of the fibre itself, because they are
just vectors tangent to a vector space. Every covector $\varphi\in\sT^\ast E$ restricted to vectors
tangent to the fibre defines an element of the space dual to the fibre. The projection $\zeta_E$
associates to a covector $\varphi$ its restriction to vectors tangent to the fibre.

Replacing the bundle $\tau$ with its dual $\pi: E^\ast\rightarrow M$ in the diagram for
$\sT E$, we get an appropriate diagram for  $\sT E^\ast$ with projections $\sT\pi$ on $\sT M$
and $\zt_{E^\ast}$ on $E^\ast$. Replacing the bundle $\tau$ with its dual $\pi$ in the diagram for
$\sT^\ast E$, we get an appropriate diagram for  $\sT^\ast E^\ast$ with projections $\pi_{E^\ast}$ on $E^\ast$
and $\zz_{E^\ast}$ on $E$. Let us notice, that the diagrams for $\sT^\ast E$ and $\sT^\ast E^\ast$
are very similar:

\medskip
\begin{minipage}[c]{0.45\linewidth}
$$\xymatrix@R+5pt{
 & \sT^\ast E^\ast\ar[dr]^{\zz_{E^*}}\ar[dl]_{\pi_{E^\ast}} & \\
E^\ast\ar[dr]^{\pi} &   & E\ar[dl]^{\tau} \\
 & M &
}$$
\end{minipage}
\hspace{0.5cm}
\begin{minipage}[c]{0.45\linewidth}
$$\xymatrix@R+5pt{
 & \sT^\ast E\ar[dr]^{\pi_E}\ar[dl]_{\zz_E} & \\
E^\ast\ar[dr]^{\pi} &  & E\ar[dl]^{\tau} \\
 & M &
}$$
\end{minipage}

\medskip
\noindent Actually, the double vector bundles $\sT^\ast E$ and $\sT^\ast E^\ast$ are canonically isomorphic.
The isomorphism
\begin{equation}\label{eq:izoR}
\mathcal{R}_E:\sT^\ast E\to\sT^* E^*
\end{equation}
is also an anti-symplectomorphism and an isomorphism of double vector bundles.
The graph of $\mathcal{R}_E$ is the Lagrangian submanifold in $(\sT^\ast
E\times\sT^\ast E^\ast, \omega_E+\omega_{E^\ast})$ generated by the evaluation of covectors and vectors
$$E\times_M E^\ast\ni(e,p)\longmapsto p(e)\in\R\,.$$
In coordinates $(x^i,y^a,p_j,\xi_a)$ in $\sT^\ast E$ and $(x^i,\xi_a,p_j,y^b)$ in $\sT^\ast E^\ast$,
the isomorphism $\mathcal{R}_E$ reads
$$\mathcal{R}_E: (x^i, y^a, p_j,\zx_b)\longmapsto (x^i,\zx_b,-p_j,y^a)\,.$$

\subsection{The classical Tulczyjew triple}
\label{sec:7}

Now we are ready to present the Lagrangian part of the Tulczyjew triple. It consists of the
map $\alpha_M$ defined in section \ref{sec:4}. The map $\alpha_M$ is an isomorphism
of double vector bundles $\sT\sT^\ast M$ and $\sT^\ast\sT M$. Both total spaces are
symplectic manifolds. The map $\alpha_M$ is also a symplectomorphism. In the following diagram
we can see the structure of $\alpha_M$ as a double vector bundle morphism:
$$\xymatrix@C-10pt@R-5pt{
 & \sT\sT^\ast M \ar[rrr]^{\alpha_M} \ar[dr]^{\sT\pi_M}\ar[ddl]_{\tau_{\sT^\ast M}}
 & & & \sT^\ast\sT M\ar[dr]^{\pi_{\sT M}}\ar[ddl]_/-10pt/{\xi} & \\
 & & \sT M\ar[rrr]^/-15pt/{id_{\sT M}}\ar[ddl]_/-10pt/{\tau_M}
 & & & \sT M \ar[ddl]_{\tau_M}\\
 \sT^\ast M\ar[rrr]^/+25pt/{id_{\sT^\ast M}}\ar[dr]^{\pi_M}
 & & & \sT^\ast M\ar[dr]^{\pi_M} & &  \\
 & M\ar[rrr]^{id_M}& & & M &
}$$
Recall that in infinitesimal mechanics, $M$ denoted the manifold of positions, $\sT M$ the manifold
of infinitesimal (kinematic) configurations, and $\sT^\ast M$ was the phase space. The constitutive set
was a subset of $\sT^\ast\sT M$ given by $\mathcal{C}=\xd L(\sT M)$, while the dynamics was defined as
$\mathcal{D}=\alpha^{-1}_M(\mathcal{C})$. W can therefore complete the diagram:
$$\xymatrix@C-10pt@R-5pt{
{ \mathcal{D}}\ar@{ (->}[r]& \sT\sT^\ast M \ar[rrr]^{\alpha_M} \ar[dr]\ar[ddl]
 & & & \sT^\ast\sT M\ar[dr]\ar[ddl] & \\
 & & \sT M\ar[rrr]\ar[ddl]
 & & & \sT M \ar[ddl]\ar@/_1pc/[ul]_{\xd L}\ar[dll]_{\cP L}\\
 \sT^\ast M\ar[rrr]\ar[dr]
 & & & \sT^\ast M\ar[dr] & &  \\
 & M\ar[rrr]& & & M &
}$$
The map $\cP L:\sT M\rightarrow \sT^\ast M$ is the {\it Legendre map} that associates momenta to velocites
and is defined as $\cP L=\xi\circ\xd L$. In coordinates $(x^i, \dot x^j)$ in $\sT M$ and
$(x^i, p_j)$ in $\sT^\ast M$, the Legendre map reads
$$\cP L(x^i, \dot x^j)=(x^i, \frac{\partial L}{\partial \dot x^j} )\,,$$
therefore $\mathcal{D}$ is given as
$$\mathcal{D}=\left\{(x^i,p_j,\dot x^k,\dot p_l):\;\; p_j=\frac{\partial L}{\partial \dot x^j},\quad \dot p_l=\frac{\partial L}{\partial x^l}\right\}\,.$$

The dynamics $\mathcal{D}\subset \sT\sT^\ast M$ is a Lagrangian submanifold with respect to the symplectic form $\xd_\sT\omega_M$,
i.e. the tangent lift of the canonical symplectic form of $\sT^\ast M$. In some cases (e.g. for hyperregular Lagrangians)
the dynamics is the image of a Hamiltonian vector field. In such a case, we can look for a Hamiltonian function that
generates the field. We observe, however, that even if $\mathcal{D}$ is not the image of a vector field, it is still worth looking for
a Hamiltonian generating object ({\it Morse family}), even if it is not as simple as just one function on $\sT^\ast M$.

In section \ref{sec:6} we have observed that two manifolds $\sT^\ast E$ and $\sT^\ast E^\ast$ are
isomorphic as double vector bundles and as symplectic manifolds. Taking $E=\sT M$, we get, according to (\ref{eq:izoR}), the canonical isomorphism
$\mathcal{R}_{TM}$ between $\sT^\ast\sT M$ and $\sT^\ast\sT^\ast M$. As a symplectic relation the isomorphism is generated by
a function $(p, v)\rightarrow \langle p,\,v\rangle$ defined on the submanifold $\sT^\ast M\times_M\sT M$ of $\sT^\ast M\times \sT M$. Following
the rules of composing symplectic relations \cite{LM}, we get that $\mathcal{R}_{TM}(\xd L(\sT M))$ is generated by a family of functions
on $\sT^\ast M$ parameterized by elements of $\sT M$,
$$\sT^\ast M\times_M \sT M\ni (p,v)\longmapsto L(v)-\langle p,\,v\rangle\in\R.$$
This most general generating object can sometimes be reduced to simpler one, but not always to just one Hamiltonian function.
The composition of double bundle morphisms $\mathcal{R}_{TM}$ and $\alpha_M$ gives the morphism
$\beta_M$, the musical isomorphism associated with the canonical symplectic structure on $\sT^\ast M$, which constitutes the Hamiltonian side of the Tulczyjew triple:
$$\xymatrix@C-10pt@R-5pt{
& \sT^\ast\sT^\ast M  \ar[dr]_{\zeta} \ar[ddl]^{\pi_{\sT^\ast M}}
 & & & \sT\sT^\ast M\ar[dr]_{\sT\pi_{M}}\ar[ddl]_/-10pt/{\tau_{\sT^\ast M}} \ar[lll]_{\beta_M}&
 { \mathcal{D}}\ar@{ (->}[l]  \\
 & & \sT M\ar[ddl]_/-10pt/{\tau_M}
 & & & \sT M \ar[ddl]_{\tau_M}\ar[lll]\\
 \sT^\ast M\ar[dr]^{\pi_M} \ar@/^1pc/[uur]^{\xd H}
 & & & \sT^\ast M\ar[dr]^{\pi_M}\ar[lll] & &  \\
 & M& & & M\ar[lll] &
}$$
It is well known that the map $\beta_M$ is associated also with the canonical symplectic structure
$\omega_M$ on $\sT^\ast M$. For any $X\in\sT\sT^\ast M$, we have $\beta_M(X)=\omega_M(\cdot, X)$.
If the Hamiltonian generating object reduces to one Hamiltonian function, then $\mathcal{D}$ is the image of the Hamiltonian vector field $X_H$ according to the formula
$$\xd H=\omega_M(\cdot, X_H).$$
The same can be written as
$$\mathcal{D}=\beta_M^{-1}(\xd H(\sT^\ast M))\,;$$
in coordinates:
$$\mathcal{D}=\left\{(x,p,\dot x,\dot p):\;\; \dot p=-\frac{\partial H}{\partial x},\quad \dot x=\frac{\partial H}{\partial p}\right\}\,.$$

The full Tulczyjew triple in mechanics is the diagram
$$\xymatrix@C-20pt@R-5pt{
&&&& { \mathcal{D}}\ar@{ (->}[d]&&&&\\
     & \sT^\ast\sT^\ast M  \ar[dr]_{} \ar[ddl]_{}
 & & & \sT\sT^\ast M \ar[rrr]^{\alpha_M} \ar[dr]_{} \ar[ddl]_{}\ar[lll]_{\beta_M}
 & & & \sT^\ast\sT M\ar[dr]^{}\ar[ddl]_/-10pt/{} & \\
 & & \sT M\ar[ddl]_/-10pt/{}
 & & & \sT M\ar[rrr]\ar[ddl]_/-10pt/{}\ar[lll]
 & & & \sT M \ar[ddl]_{}\ar[ddl]_{}\ar@/_1pc/[ul]_{\xd L}\\
\sT^\ast M\ar[dr]^{}\ar@/^1pc/[uur]^{\xd H}
 & & & \sT^\ast M\ar[rrr]\ar[dr]^{}\ar[lll]
 & & & \sT^\ast M\ar[dr]^{} & &  \\
 & M & & & M\ar[rrr]\ar[lll]& & & M &
}$$
%
Using the structure encoded in the Tulczyjew triple, we can describe more complicated mechanical systems than those
known in the traditional Lagrangian and Hamiltonian mechanics. In geometrical optics, for example, there are systems
for which we need more general generating object on the Lagrangian side. In relativistic mechanics, we need generating
families on the Hamiltonian side. The above diagram shows also that, from the mathematical point of view, Hamiltonian and
Lagrangian mechanics are equivalent only if we agree to use these more general generating objects. We should however
keep in mind that Lagrangian mechanics has variational origins and comes from Lagrangian mechanics for a finite time interval. We understand the Hamiltonian mechanics as an alternative way of generating the dynamics of a system: the image of the dynamics by the map $\beta_M$ is a lagrangian submanifold of the cotangent bundle $\sT^\ast\sT^\ast M$, therefore we can consider its generating objects. This works only in the infinitesimal formulation, so that Hamiltonian formalism is genuinely infinitesimal. For finite time interval, the only isomorphism of $\mathbb{P}Q$ with a cotangent bundle is $\alpha$, associated with Lagrangian mechanics, and no Hamiltonian description is available.
\subsection{Mechanics on algebroids}
\label{sec:7a}
We can generalize the classical Tulczyjew triple {\it mutatis mutandis} to a mechanics on algebroids \cite{GG,GGU,LMM}. The starting point is the diagram in which a vector bundle $E$ over $M$ replaces $\sT M$ and $\za$ and $\zb$ are replaced by morphisms in the reverse directions:
$$\xymatrix@C-10pt@R-5pt{
&&&& { \mathcal{D}}\ar@{ (->}[d]&&&&\\
 &\sT^\ast E^\ast \ar[rrr]^{\ze\circ \mathcal{R}_E^{-1}}
\ar[ddl] \ar[dr]
 &  &  & \sT E^\ast \ar[ddl] \ar[dr]
 &  &  & \sT^\ast E \ar[ddl] \ar[dr]
\ar[lll]_{\varepsilon}
 & \\
 & & E \ar[rrr]^/-20pt/{\zr}\ar[ddl]
 & & & \sT M\ar[ddl]
 & & & E\ar[lll]_/+20pt/{\zr}\ar[ddl]\ar[ddl]_{}\ar[ddl]_{}\ar@/_1pc/[ul]_{\xd L}
 \\
 E^\ast\ar[rrr]^/-20pt/{id} \ar[dr]\ar[dr]^{}\ar@/^1pc/[uur]^{\xd H}
 & & & E^\ast\ar[dr]
 & & & E^\ast\ar[dr]\ar[lll]_/-20pt/{id}
 & & \\
 & M\ar[rrr]^{id}
 & & & M & & & M\ar[lll]_{id} &
}$$
Thus, $\ze:\sT^* E\to\sT E^*$ represents the structure of an algebroid with the {\it anchor map} $\zr:E\to\sT M$. The connection with the standard definition of an algebroid by means of a bracket of sections can be found in
\cite{GU,Gr2}. The construction of dynamics is the same:
the left-hand side is Hamiltonian with Hamiltonians being functions $H:E^*\to\R$,
the right-hand side is Lagrangian with Lagrangians being functions $L:E\to\R$, and the phase dynamics lives in the middle.
Note finally that the above formalisms can still be generalized to
include constraints (cf. \cite{GG1}) and that a rigorous optimal
control theory on Lie algebroids can be developed as well
\cite{CM,GJ}.

\section{The geometry of classical field theories}
\label{sec:8}

Following the statical ideas we can go into classical field theory. Like in the case of mechanics,
we start with finite domain formulation to find geometric models for physical quantities. We need
configurations, processes, cost functions, and constitutive sets. In the following we will not go
into the details of all these constructions. Since general rules are already known, we can just list main
results.

\subsection{Classical fields for bounded domains}
\label{sec:9}

Configurations in $Q$ (fields) are smooth sections $q:D\to E$ of a locally trivial fibration $\zz:E\to M$
over a manifold $M$ of dimension $m$, supported on compact discs $D\subset M$ with smooth boundary
$\pa D$. We will use the coordinates $(x^i,y^a)$ in $E$. Parameterized processes $s\mapsto q_s$ come from vertical homotopies
$\chi:\R\ti M\supset I\ti \mathcal{O}\to E$, $D\subset\mathcal{O}$, so that infinitesimal processes (tangent vectors) in convenient representation
are vertical vector fields $\zd q$ over $D$, $\zd q:D\to\sV E$ (Fig.\ref{fig:14}).
\begin{pict}[ht]
\begin{floatrow}[1]
\floatbox{pict}[0.4\textwidth]
{\caption{Infinitesimal process}\label{fig:14}}
{\centering\includegraphics[width=.3\textwidth]{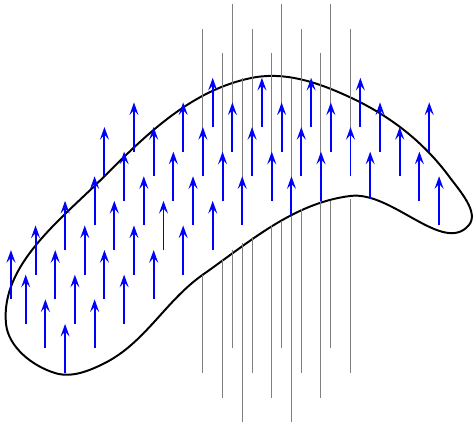}}
\end{floatrow}
\end{pict}
Functions are associated with first-order Lagrangians, i.e. bundle maps $L:\sJ^1E\to\zW^m$, where we denote $\zW^k=\wedge^k\sT^*M$,
$$S(q)=\int_DL(j^1q)\,.$$
Covectors are equivalence classes of functions. It was very easy to guess convenient representatives for tangent vectors. To do the same
for covectors, we need some calculations. According to Stokes theorem,
\begin{equation}\label{eq:stokes}
\frac{\xd }{\xd s}_{|s=0}S(q_s)=
\int_D\langle\mathcal{E}L\circ j^2 q,\zd q\rangle+\int_{\pa D} \langle\,\cP L\circ j^1q,\delta q\,\rangle\,.
\end{equation}
Here
\begin{align}
\mathcal{E}L: \sJ^2E&\longrightarrow \sV^* E\otimes_E\zW^m \\
\intertext{is the \emph{Euler-Lagrange operator} and}
\cP L=\zx\circ\xdv L:J^1E&\longrightarrow\sV^* E\otimes_E\zW^{m-1}\\
\intertext{is the \emph{Legendre map}, where $\zx$ is a certain canonical map}
\zx:\sV^*\sJ^1E\ot_E\zW^m&\longrightarrow\sV^* E\ot_E\zW^{m-1}\,.
\end{align}
The map $\zx$ is analogous to the second projection $\zz_E$ in the structure of the double vector bundle $\sT^\ast E$
(see section \ref{sec:6}), although $\sV^*\sJ^1E\ot_E\zW^m$ is not a double vector bundle, but double
vector-affine bundle \cite{GRU}.
It follows that covectors are represented by pairs of sections $(f,p)$, where
\begin{align}
f:D&\longrightarrow \sV^+E=\sV^* E\otimes_E\zW^m\,, \\
p:\pa D&\longrightarrow\cP E=\sV^* E\otimes_E\zW^{m-1}\,.
\end{align}
We have found additionally that the phase space (in mechanics the space of momenta) is $\cP E$. Since
we are again interested in differential equations describing the fields and phase sections, we pass
immediately to infinitesimal formulation. The section $f$ plays the role of the source of a field.
In the infinitesimal approach we shall put sources equal to zero for simplicity, however in principle
they can be added to the picture.

\subsection{Classical fields: Infinitesimal version}
\label{sec:10}

Infinitesimal formulation arises when we, informally speaking, integrate over an
infinitesimal domain of $M$. In such a case configurations are elements of $Q=\sJ^1E$. In
$\sJ^1 E$, we will use coordinates $(x^i, y^a, y^c_k)$ coming from coordinates in $E$.
Now, the space of configurations is again a manifold, however we have to use density valued
functions $L:\sJ^1E\to\zW^m$ instead of real valued functions.
We assume that $M$
is oriented and we can identify densities with $m$-forms.
Since $Q$ is a manifold, we know what
tangent vectors are, but we keep in mind that in finite domain formulation we used only
vertical homotopies. Starting with a vertical homotopy $\chi:I\times \mathcal{O}\rightarrow E$
and taking first the infinitesimal part with respect to $M$ ($s\mapsto\sj^1\chi(s,x)$) and then
tangent vector in the vertical direction ($\delta\sj^1\chi(0,x)$), we get $\sT Q=\sV\sJ^1 E$.
Convenient representatives of tangent vectors in finite domain formulations were obtained by taking
the tangent vector in vertical direction first ($\delta\chi(0,\cdot)$). We now need an infinitesimal
part of that vertical vector field, i.e. $\sj^1\delta\chi(0,x)$. The correspondence between vectors
tangent to $Q$ and their convenient representatives is now expressed as an isomorphism $\kappa:
\sV\sJ^1E\rightarrow \sJ^1\sV E$ of double vector-affine bundles,
$$\xymatrix@C-20pt@R-8pt{
 & \sV\sJ^1 E\simeq\sJ^1\sV E\ar[dr]\ar[dl]_{} & \\
\sJ^1E\ar[dr]^{} &   & \sV E\ar[dl]^{} \\
 & E &
}$$

\medskip
Covectors on $Q$ come from classifying functions with respect to vertical curves in $Q$.
This means that we get $\sV^\ast\sJ^1E\otimes_{\sJ^1E}\zW^m$. Since the notation here
is becoming heavy, we choose another symbol for the space of covectors. From now on
$\sV^+\sJ^1E$ will denote $\sV^\ast\sJ^1E\otimes_{\sJ^1E}\zW^m$. The infinitesimal
version of the application of the Stokes theorem (\ref{eq:stokes}) reads
\begin{equation}\label{eq:infstokes}
\langle\xd L, \delta\sj^1\chi\rangle=\langle\mathcal{E}L(\sj^2\chi),\delta\chi\rangle+
\xd(\langle\mathcal{P}L(\sj^1\chi), \delta\chi\rangle).
\end{equation}
Since the evaluation of a section of $\cP E\rightarrow M$ with a section of $\sV E$ is
an $(m-1)$-form on $M$, we can differentiate it and evaluate at a point $x\in M$. The result
depends on the first jet of the section of $\cP E\rightarrow M$ and a first jet of vertical
vector field. In this way, we have obtained a bilinear evaluation,
$$
\langle\!\langle\cdot,\cdot\rangle\!\rangle: \sJ^1\mathcal{P}E\times_{\sJ^1 E}\sJ^1\sV E\longrightarrow \Omega^m\,,
$$
defined on $\sj^1 p(x_0)$ and $\sj^1\delta\sigma(x_0)$ by the formula
\begin{equation}\label{eq:evalfield}
\langle\!\langle\,\sj^1 p(x_0),\sj^1\delta\sigma(x_0)\,\rangle\!\rangle=
\xd \langle p,\delta\sigma\rangle(x_0)\,.
\end{equation}
The identification of $\Hom(\sV\sJ^1E,\zW^m)$ with $\sV^+\sJ^1E$ defines the map
\begin{equation}\label{eq:alphafield}
\alpha:\sJ^1\cP E \longrightarrow \sV^+\sJ^1E
\end{equation}
which is dual to $\kappa$.
In the adapted coordinates $(x^i,y^a,p^j_b,y^c_k,p^l_{dm})$  in
$\sJ^1\cP E$ and $(x^i,y^a,y^c_k,\pi_d, \pi^l_e)$ in $\sV^+\sJ^1 E$,  we have
$$\za(x^i,y^a,p^j_b,y^c_k,p^l_{dm})=(x^i,y^a,y^c_k,\sum_lp^l_{dl}, p^j_b)\,.$$
We have used here the same letters $\kappa$ and $\alpha$ that appeared already in the context
of mechanics (\ref{sec:3a}). In mechanics for finite time interval they denoted te correspondence between
vectors and covectors on configurations and their convenient representations. In field theory
they play exactly the same role.

\subsection{Tulczyjew triple: Lagrangian side for field theory}
\label{sec:11}

The map $\alpha$ constitutes the Lagrangian side of the Tulczyjew triple for first
order field theory. The Lagrangian side can be written in a form of a diagram
$$\xymatrix@C-15pt@R-5pt{
{ \mathcal{D}}\ar@{ (->}[r]& \sJ^1\cP E \ar[rrr]^{\alpha} \ar[dr]_{} \ar[ddl]_{}
 & & & \sV^+ \sJ^1E\ar[dr]^{}\ar[ddl]_/-10pt/{} & \\
 & & \sJ^1E\ar[rrr]\ar[ddl]_/-10pt/{}
 & & & \sJ^1E \ar[ddl]_{}\ar@/_1pc/[ul]_{\xdv L}\ar[dll]_{\cP L}\\
 \cP E\ar[rrr]\ar[dr]^{}
 & & & \cP E\ar[dr]^{} & &  \\
 & E\ar[rrr]& & & E &
}$$
According to the rules we developed while analyzing the mechanical triple, the dynamics of the field consists of
the convenient representatives of elements of the constitutive set, i.e.
\begin{equation}\label{eq:dynamicfield}
\mathcal{D}=\alpha^{-1}(\xdv L(\sJ^1E)).
\end{equation}
There is also the Legendre map,
$$\cP L:\sJ^1E\rightarrow\cP E, \quad \cP L=\xi\circ\xdv L\,,$$
that associates phase elements to configuration elements. In coordinates, the dynamics reads
$$\mathcal{D}=\left\{(x^i,y^a,p^j_b,y^c_k,p^l_{dm}):\;\; p^j_b=\frac{\partial
L}{\partial y^b_j},\quad \sum_lp^l_{dl}=\frac{\partial L}{\partial y^d}\right\}\,,$$
while the Legendre map is
$$\cP L(x^i,y^a,y^b_j)= \left(x^i,y^a,\frac{\partial L}{\partial y^b_j}\right)\,.$$
We get also the Euler-Lagrange equations
$$\frac{\pa L}{\pa y^a}=\frac{\pa}{\pa x^i}\frac{\pa L}{\pa y^a_i}\,.$$

The manifolds $\sJ^1\cP E$ and $\sV^+ \sJ^1E$ are both double vector-affine bundles with affine structure over
$\cP E$ and linear structure over $\sJ^1 E$. They both carry some sort of a symplectic structure. Every
fibre of $\sV^+ \sJ^1E$ over $M$ is a manifold equipped with a symplectic form with values in $\zW^m$.
Every fibre of $\sJ^1\cP E$ over $M$ is a manifold equipped with presymplectic form with values in $\zW^m$.
The map $\alpha$ is a morphism of double bundle structures and symplectic structures. This time, however, it is not
an isomorphism. It is possible to reduce the space $\sJ^1\cP E$ to get an isomorphism, but then we loose the natural
interpretation of the dynamics as a first order partial differential equation, because it is no more a subset of
a first jet bundle.

\subsection{Tulczyjew triple: Hamiltonian side for field theory}
\label{sec:12}

Defining the Hamiltonian side of the triple in field theory means looking for another generating object
for the dynamics. In mechanics, we could use two structures: one was the duality between momenta and
velocities that allowed us to define the symplectic relation $\mathcal{R}_{TM}$ (see (\ref{eq:izoR})), the other
was the canonical structure on the phase space. Both ways can be followed in field theory, however not
in an easy way. We shall concentrate on the analog of the duality between momenta and
velocities. First of all, let us notice that there is no duality between the configuration space
$\sJ^1 E$ and the phase space $\cP E$. The phase elements are naturally evaluated on virtual displacements,
not on configurations. Actually, in mechanics we have the same: from the construction of momenta we get that
they are to be evaluated on virtual displacements. In mechanics, virtual displacements are represented by the same
geometrical objects as infinitesimal configurations, therefore we can write $\langle p,\dot x\rangle$ as well
as $\langle p,\delta x\rangle$. It is not the case in field theory. To find objects dual to
infinitesimal configurations, we have to use affine geometry.

Let us fix a point $x$ in $M$. Having in mind the contents of section \ref{sec:1} we can take
the affine bundle $A=\sJ^1_xE$ over $N=E_x$, and the vector space $Y=\zW^m_x$.
Let $\sJ^\dag_xE=A^\dag(Y)=\Aff(A,Y)$ be the affine-dual bundle which is an AV-bundle over $\sT_x M\ot\sV^*_xE\ot\zW^m_x\simeq \cP_xE$,
$$\zvy_x:\sJ^\dag_xE\to\cP_xE\,,\quad (y^a,p^j_b,r)\mapsto(y^a,p^j_b)\,.$$
Let $\sP\sJ^\dag_x E$ be the corresponding affine phase bundle of affine differentials $\uxd H_x$ of
sections $H_x:\cP_xE\to\sJ^\dag_xE$ (see \ref{def:phase}). Collecting the affine phase bundles $\sP\sJ^\dag_x E$ point by
point in $M$, we obtain the affine phase bundle  $\sP\sJ^\dag E$ which is the bundle of `vertical affine
differentials' $\uxd^v H$ of sections $H$ of the bundle $\zvy:\sJ^\dag E\to\cP E$,
$$\sP\zvy:\sP\sJ^\dag E\to\cP E\,,$$
In natural coordinates in $\sP\sJ^\dag E$, the projection reads
$$\quad(x^i,y^a,p^j_b,p_c,y^d_l)\mapsto(x^i,y^a,p^j_b)\,.$$
The bundle $\sP\sJ^\dag E$ is actually a double affine-vector bundle  isomorphic canonically with $\sV^+\sJ^1E$,
$$\xymatrix@C-15pt@R-8pt{
 & \sV^+\sJ^1E \ar[rrr]^{\mathcal{R}} \ar[dr]^{}
 \ar[ddl]_{\zx}
 & & & \sP\sJ^\dag E\ar[dr]^{}\ar[ddl]^/10pt/{\sP\zvy}
 & \\
 & & \sJ^1E\ar[rrr]^/-20pt/{}\ar[ddl]_/-20pt/{}
 & & & \sJ^1E \ar[ddl]_{}\\
 \cP E\ar[rrr]^/-20pt/{id}\ar[dr]^{}
 & & & \cP E\ar[dr]^{} & &  \\
 & E\ar[rrr]^{id}& & & E &
}
$$
The map $\mathcal{R}$ is generated analogously to $\mathcal{R}_E$ by evaluation between
elements of $\sJ^1 E$ and $\sJ^\dag E$ over $E$. The construction involves some affine
geometry and symplectic reduction. The details can be found in \cite{G}.
Composing $\za$ with $\mathcal{R}$, we get a map $\zb=\mathcal{R}\circ\za$ constituting the
Hamiltonian side of the triple
$$\zb:\sJ^1\cP E\to \sP\sJ^\dag E\,.$$
In adapted coordinates,
$$\zb(x^i,y^a,p^j_b,y^c_k,p^l_{dm})=(x^i,y^a,p^j_b,-\sum_lp^l_{dl},y^c_k)\,.$$

The Hamiltonian side of the Tulczyjew triple for field theory can be written in a form of a diagram
$$\xymatrix@C-10pt@R-5pt{
& \sP\sJ^\dag E  \ar[dr]_{} \ar[ddl]^{\sP\zvy}
 & & & \sJ^1\cP E\ar[dr]^{}\ar[ddl]_/-10pt/{} \ar[lll]_{\beta}& { \mathcal{D}}\ar@{ (->}[l] \\
 & & \sJ^1E\ar[ddl]_/-10pt/{}
 & & & \sJ^1E \ar[ddl]_{}\ar[lll]\\
 \cP E\ar[dr]^{} \ar@/^1pc/[uur]^{\uxd^v H}
 & & & \cP E\ar[dr]^{}\ar[lll] & &  \\
 & E& & & E\ar[lll] &
}$$
Hamiltonians are sections of the one-dimensional affine bundle $\zvy:\sJ^\dag E\to\cP E$.
The dynamics is generated by means of $\beta$ by
$$\mathcal{D}=\beta^{-1}(\uxd^v H(\cP E))\,.$$
In adapted coordinates, we get
$$\mathcal{D}=\left\{(x^i,y^a,p^j_b,y^c_k,p^l_{dm}):\;\; \sum_l p^l_{dl}=-\frac{\partial H}{\partial y^d},\quad y^c_k=\frac{\partial H}{\partial p^k_c}\right\}\,.$$
Of course, on the Hamiltonian side we can encounter all the problems we have in mechanics, concerning the fact that
the dynamics is not always generated by one section. We can always get a generating family of sections adding generating
objects of the constitutive set and the relation $\mathcal{R}$.

\subsection{The Tulczyjew triple for field theory}
\label{sec:13}

The complete Tulczyjew triple for first order field theory has the form of the following diagram.
$$
\xymatrix@C-15pt{
&&&& { \mathcal{D}}\ar@{ (->}[d]&&&&\\
 & \sP\sJ^\dag E\ar[dl]_{}\ar[ddr]^/-10pt/{} & & &
 \sJ^1\mathcal{P}E\ar[lll]_\beta\ar[rrr]^\alpha\ar[dl]_{}\ar[ddr]^/-10pt/{}& & &
 \sV^+\sJ^1E \ar[dl]\ar[ddr]^/-10pt/{}& \\
{\mathcal{P}E\ }\ar[ddr]^/-10pt/{}\ar@/^1pc/[ur]^{\uxd^v H} & & &
{\mathcal{P}E\ }\ar[rrr]\ar[lll]\ar[ddr]^/-10pt/{} & & & {\ \mathcal{P}}E\ar[ddr]^/-10pt/{} & & \\
    & & \sJ^1 E\ar[dl]_{} & & & \sJ^1 E\ar[rrr]\ar[lll]\ar[dl]_{} & & & \sJ^1 E\ar[dl]_{}\ar@/_1pc/[uul]_{\xdv L} \\
 & E &  & & E\ar[rrr]\ar[lll] & & & E &
}
$$

\medskip\noindent
All the three double bundles are double vector-affine bundles. The structure over $\cP E$ is affine, while
the structure over $\sJ^1 E$ is linear. The Lagrangian and Hamiltonian bundles are isomorphic.
They are both, fiber by fiber over $M$, equipped with canonical symplectic forms with values in $\zW^m$.
The canonical structure of the phase bundle is the tautological form $\vartheta_\cP$ which, in adapted coordinates, reads
$$\vartheta_\cP=p^i_a\xd y^a\otimes\eta_i,$$
where $\eta_i=\imath(\partial_i)\eta$ with $\eta=\xd x^1\wedge\cdots\wedge\xd x^m$. The tautological form, differentiated
vertically, gives the form
$$\omega_\cP= (\xd p^i_a\wedge\xd y^a)\otimes\eta_i$$
with values in $\zW^{m-1}$. The latter, lifted to $\sJ^1\cP E$, is,
fiber by fiber over $M$, a presymplectic form with values in $\zW^m$.

We have observed that Hamiltonians are sections of the one-dimensional affine bundle $\sJ^\dag E\rightarrow \cP E$.
The space $\sJ^\dag E$, i.e. the space of the values of Hamiltonians can be identified with
a subspace of $m$-forms on $E$ with the property that they vanish, while evaluated on two vertical vectors \cite{V}.
We denote this space by $\wedge^m_1\sT^\ast E$. To see the identification, we have to be able to evaluate an element of
$\wedge^m_1\sT^\ast E$ on the first jet of a section of $\zz$. Let us use coordinates to simplify
the presentation. En element $\varphi$ of $\wedge^m_1\sT^\ast E$ can be written locally as
$$\varphi=A\,\eta+B^i_a dy^a\wedge\eta_i\,.$$
Note that here we have used a different $dy^a$ than previously. The difference is that
$\xd y^a$ is an element of $\sV^\ast E$, while $dy^a$ is an element of $\sT^\ast E$. Using the first
jet given in coordinates by $(x^i, y^a, y^b_j)$, we can split the space $\sT E$ at point $(x^i,y^a)$
into vertical vectors $\sV E$ and horizontal vectors. Vertical space is spanned by vectors $(\partial_a)$,
while horizontal by $(\partial_i+y^a_i\partial_a)$. The dual space $\sT^\ast E$ is also split. The anihilator
of the space of vertical vectors is spanned by $(\xd x^i)$, and the anihilator of the space of horizontal
vectors is spanned by $(dy^a-y^a_i\xd x^i)$. We can therefore identify vertical differentials $\xd y^a$
with $dy^a-y^a_i\xd x^i$. Let us look at elements of $\wedge^m_1\sT^\ast E$ in the basis induced by
the jet:
\begin{equation}\label{eq:split}
\varphi=A\,\eta+B^i_a dy^a\wedge\eta_i=
A\,\eta+B^i_a (\xd y^a+y^a_j\xd x^j)\wedge\eta_i=
(A+B^j_ay^a_j)\eta+B^i_a\xd y^a\wedge\eta_i\,.
\end{equation}
We see in (\ref{eq:split}) that, using the jet, we can split also the space $\wedge^m_1\sT^\ast E$ into purely
horizontal forms and forms that have one vertical factor. The value of an affine map corresponding to
$\varphi$ is just the horizontal part of the form $\varphi$ under the splitting induced by the jet.
The part with one vertical factor can be written now as $B^i_a\xd y^a\otimes\eta_i$ and identified
with the projection of an element of $\wedge^m_1\sT^\ast E$ on $\cP E$.

\subsection{Example: Tulczyjew triple for time-dependent systems}
\label{sec:14}

Our first example will be the Tulczyjew triple for a time dependent system for a fixed observer, i.e.
when we can write the space of positions in the form of cartesian product with time.
We put here $\zz:E=Q\ti\R\to\R=M$ and get the following identifications:
\begin{align*}
\sJ^1E&\simeq\sT Q\ti\R\,,\\
\cP E&\simeq \sT^*Q\ti\R\,,\\
\sV^+\sJ^1E&\simeq \sT^*\sT Q\ti\R\,,\\
\sJ^1\cP E&\simeq \sT\sT^* Q\ti\R\,,\\
\sP\sJ^\dag E&\simeq \sT^*\sT^* Q\ti\R\,.
\end{align*}
Thus, the Tulczyjew triple takes the form
$$\xymatrix@C-35pt@R-5pt{
 &&&& { \mathcal{D}}\ar@{ (->}[d]&&&&\\
 & \sT^\ast\sT^\ast Q\ti\mathbb R \ar[dr]^{}
 \ar[ddl]_{}
 & & & \sT\sT^\ast Q\ti\mathbb R\ar[lll]_{{\beta}}\ar[rrr]^{\alpha}\ar[dr]^{}\ar[ddl]_/-20pt/{}
 &  &  & \sT^\ast\sT Q\ti\mathbb R \ar[ddl]_/-25pt/{} \ar[dr]^{}
 & \\
 & & \sT Q\ti\mathbb R\ar[ddl]
 & & & \sT Q\ti\mathbb R \ar[lll]_/+10pt/{} \ar[rrr]^/-10pt/{}\ar[ddl]_{}
& & & \sT Q\ti\mathbb R\ar[ddl]_{}\ar@/_1pc/[ul]_/-5pt/{\xdv L}
 \\
 \sT^\ast Q\ti\mathbb R\ar[dr]^{}\ar@/^1pc/[uur]^{\uxd^v H}
 & & & \sT^\ast Q\ti\mathbb R\ar[rrr]^{}\ar[dr]^{}\ar[lll]_{} & & & \sT^\ast Q\ti\mathbb R
 \ar[dr]^{}
 & &  \\
 & Q\ti\mathbb R& & & Q\ti\mathbb R\ar[lll]_{id}\ar[rrr]^{} & & & Q\ti\mathbb R&
}
$$

\subsection{Example: Scalar fields}
\label{sec:15}
The theory of a scalar field is based on the fibration $E=M\times\R\rightarrow M$. On the Lagrangian side, we get the following identifications:
$$\begin{aligned}
\sJ^1 E & \simeq  \R\times\sT^\ast M\,, &  (\varphi,f)\,,\\
\mathcal{P}E &\simeq \R\times\Omega^{m-1}\,,&  (\varphi,p)\,, \\
\sV^+\sJ^1 E& \simeq  \R\times\sT^\ast M\times_M\Omega^{m}\times_M\Omega^{m-1}\,, & (\varphi,f,a,p)\,, \\
\sJ^1\mathcal{P} & \simeq   \R\times\sT^\ast M\times_M\sJ^1\Omega^{m-1}\,, & (\varphi, f, \sj^1 p)\,.
\end{aligned}$$
The map
$\alpha: \sJ^1\mathcal{P}\longrightarrow \sV^+\sJ^1 E$ reads
$$\alpha(\varphi, f, \sj^1 p)=(\varphi,f,\xd p(x), p(x)),$$
where $x\mapsto p(x)$ is any representative of $\sj^1 p$.
Let $g$ be a metric tensor on $M$. Denote with $G:\sT M\rightarrow \sT^\ast M$ the corresponding isomorphism, with
$\omega$ -- the volume form associated with the metrics, and with $\star f=G^{-1}(f)\,\lrcorner\,\omega$ -- the Hodge operator. For the Lagrangian
$$L(\varphi,f)=\frac12 f\wedge\star f\,,$$
we get
$$\xd^{\sv}L(\varphi,f)=(\varphi,\; f,\; 0,\; \star f)\quad{
\in\R\times\sT^\ast M\times_M\Omega^{m}\times_M\Omega^{m-1}}.$$
A section $M\ni x\longmapsto (\varphi(x), p(x))\in\cP E$ is a solution of
the Lagrange equations if
$$\sj^1(\varphi,p)(x)\in\alpha^{-1}(\xd^\sv L(\varphi(x),\xd \varphi(x)))\,,$$
i.e.
\begin{align*}
\xd \varphi(x)&=f\,, \\
p&=\star f\,, \\
\xd p&=0\,.\end{align*}
The corresponding Euler-Lagrange equation is therefore
$$\xd \star\xd \varphi=0,\quad\text{i.e.}\quad \Delta \varphi=0\,.$$

Since $\sJ^1E=\R\times\sT^\ast M\rightarrow M\times\R=E$ is a vector bundle, the Hamiltonian
side is simplified.
The fibre of the bundle $\sJ^1E\rightarrow E$ over $(\varphi, x)$ is equal
to $\sT_x^\ast M$, so any affine map $A_{\varphi,x}:\sT_x^\ast M\rightarrow \Omega_x^m$ on the fibre takes the form
$$A_{\varphi,x}(f)=f\wedge p+a\,,$$
where $p\in\Omega_x^{m-1}$ and $a\in\Omega_x^m$.

We have therefore
$$\begin{aligned}
\sJ^\dag E & \simeq  \R\times\Omega^{m-1}\times_M\Omega^{m}\,,  & (\varphi, p, a)\,, \\
\sP \sJ^\dag E & \simeq  \R\times\Omega^{m-1}\times_M\sT^\ast M\times_M \Omega^m\,, &  (\varphi,p, f, a)\,. \\
\end{aligned}$$
The map $\beta: \sJ^1\mathcal{P}\longrightarrow \sP\sJ^\dag E$ reads
$$\beta(\varphi, f, \sj^1 p)=(\varphi,p(x),f,\xd p(x)),$$
where $x\mapsto p(x)$ is any representative of $\sj^1 p$.
Since the bundle $\theta: \sJ^\dag E\rightarrow \mathcal{P}$ is trivial, Hamiltonians are maps $H:\mathcal{P}\rightarrow
\Omega^m$. For the Hamiltonian
$$H(\varphi, p)=\frac12p\wedge\star p\,,$$
we get
$$\xd^{\sv}H(\varphi, p)=(\varphi, p, \star p, 0).$$
The Hamilton equations for a section $M\ni x\longmapsto (\varphi(x), p(x))$ read
\begin{align}
\xd \varphi(x)&=\star p, \\
\xd p&=0.\end{align}
The above equations lead to the following equation for hr field $x\mapsto \varphi(x)$:
$$\xd\star\xd\varphi=0,\quad\text{i.e.}\quad \Delta \varphi=0.$$

\subsection{Example: Vector fields}
\label{sec:16}

Let us suppose that the bundle $\zz: E\rightarrow M$ is a vector bundle. In such a case, the bundle $\sJ^1 E\rightarrow M$ is also a vector bundle with a distinguished subbundle
$W=\{\sj^1\sigma(x):\; \sigma(x)=0_x\}$. The space $\sT_{0_x}E$ is a direct sum of
the space of vertical vectors $\sV_{0_x}E\simeq E_x$ and the space of vectors tangent
to the zero-section which can be identified with $\sT_xM$. It follows that
$W\simeq \sT^\ast M\otimes E$. The projection $\sJ^1E\rightarrow E$ coincides
with the projection $\sJ^1 E\rightarrow (\sJ^1 E\slash W)\simeq E$. Let us denote
with $\sJ^\ast E\rightarrow M$ the bundle dual to $\sJ^1 E\rightarrow M$. There is
a projection $\sJ^\ast E\rightarrow W^\ast\simeq \sT M\otimes E^\ast$. We have the identifications
\begin{align*}
\sV^+\sJ^1 E&\simeq \sJ^1E\times_M(\sJ^\ast E\otimes\Omega^m)\,, \\
\cP E&\simeq E\times_M (W^\ast\otimes\Omega^m)\,, \\
\sJ^1\cP&\simeq \sJ^1E\times_M \sJ^1(W^\ast\otimes\Omega^m)\,.
\end{align*}
The map
$$\alpha: \sJ^1E\times_M \sJ^1(W^\ast\otimes\Omega^m) \longrightarrow \sJ^1E\times_M\sJ^\ast E\otimes\Omega^m $$
separates into two maps $\alpha=\alpha_1\times\alpha_2$. The first factor is the identity on $\sJ^1E$,
and the second is a bundle morphism
$$\xymatrix{ \sJ^1(W^\ast\otimes\Omega^m)\ar[r]^{\alpha_2}\ar[d] & \sJ^\ast E\otimes\Omega^m \ar[d] \\
W^\ast\otimes\Omega^m\ar[r]^{=}&  W^\ast\otimes\Omega^m}\,.$$
Starting from coordinates $(x^i, y^a)$ linear in fibres of $\zeta$, we get coordinates
$(x^i, \varphi_a, \varphi^j_b)$ in $\sJ^\ast E\otimes\Omega^m$ linear in fibres over $M$
and $(x^i, p^i_a, p^j_{bl})$ in $\sJ^1(W^\ast\otimes\Omega^m)$. The map $\alpha_2$ reads
$$\alpha_2(x^i, p^j_a, p^k_{bl})=(x^i, \sum_j p^j_{aj}, p^k_b).$$

On the Hamiltonian side, in view of theorem \ref{th:1} in its version for vector spaces, there is a canonical identification,
$$\sP\sJ^\dag E\simeq\sJ^1 E\times_M\sJ^\ast E \otimes\Omega^m\,,$$
with the two projections: $pr_1$ on $\sJ^1 E$ and $\xi$ on $E\times_M W^\ast\otimes \Omega^m$.
Out of coordinates $(x^i, y^a, p^i_b, r)$ in $\sJ^\dag E$, we get coordinates
$(x^i, y^a, p^j_b, \pi_b, \pi^c_k)$ in $\sP\sJ^\dag E$ with projections
\begin{align*}
(x^i, y^a, p^j_b, \pi_b, \pi^c_k)& \longmapsto (x^i, y^a, \pi^c_k)\in\sJ^1 E \\
\intertext{and}
(x^i, y^a, p^j_b, \pi_b, \pi^c_k)& \longmapsto (x^i, y^a, p^j_b)\in E\times_M W^\ast\otimes\Omega^m.
\end{align*}
The map $\beta$ reads
\begin{align*}
\beta: \sJ^1E\times_M \sJ^1(W^\ast\otimes\Omega^m)&\longrightarrow \sJ^1 E\times_M\sJ^\ast E \otimes\Omega^m\,;\\
\intertext{in coordinates:}
(x^i y^a, y^b_j, p^k_c, p^l_{dm})&\longmapsto (x^i, y^a, p^j_b, -\sum_kp^k_{ck}, y^b_j).
\end{align*}
The Lagrangian and Hamiltonian spaces are, as usual, isomorphic. Here, it is even more visible, because of theorem \ref{th:1}. However, we have to remember that, on the Lagrangian side, the dynamics
is generated out of the map $L:\sJ^1E\rightarrow\Omega^m$, and on the Hamiltonian side out of the section
$H:E\times W^\ast\otimes\Omega^m\rightarrow \sJ^\dag E$. In some sense, the Lagrangian side is associated with the projection $pr_1$ on $\sJ^1E$, while the Hamiltonian side with the projection $\xi$ on $E\times W^\ast\otimes\Omega^m$.

\subsection{Example: Electromagnetics}
\label{sec:20}
Now let us check how Electromagnetics, i.e the true physical theory, fits into the general scheme.
In Electrodynamics, fields (electromagnetic potentials, $A$),
are one forms on the four
dimensional manifold $M$ equipped with a metrics with Lorenz signature, but we can write with $M$ of arbitrary dimension $m>1$. In our model,
$E=\sT^\ast M$  and $\zz=\pi_M$. The symbols $\vee$
and $\wedge$ denote the symmetrized and antisymmetrized tensor product, respectively. The canonical density associated
to the metric is $\omega$, while $\omega_M$ stands, as usual, for the canonical symplectic form on $\sT^\ast M$.

Let us take a closer look at the structure of the vector space
$\sJ^1_x\sT^\ast M$ for a fixed $x\in M$. Like in the general case of a vector field, the space $\sJ^1_x\sT^\ast M$
is a vector space with the distinguished subspace $W_x$ of jets of one-forms on $M$ which take the value $0$
at $x$. It follows from the general considerations concerning vector fields that $W_x\simeq \sT^\ast_xM\otimes\sT^\ast_xM$.
Using the canonical splitting of two tensors into symmetric and antisymmetric parts, we get that
$W_x\simeq \vee^2\sT^\ast_xM\oplus\wedge^2\sT^\ast_x M$.
In $\sJ^1_x\sT^\ast M$ there is another
vector subspace $S_x$ of jets of closed forms. Since everything is local here, we can consider them as
jets of differentials of local functions on $M$. It is easy to see that
$\vee^2\sT^\ast_xM=W_x\cap S_x$ and $\sJ^1_x\sT^\ast M=W_x\oplus S_x$. Moreover, there is an isomorphism
$\sJ^1_x\sT^\ast M\slash S_x\simeq \wedge^2\sT^\ast_x M$.

Note also canonical maps:
$$\zg:\sJ^1\sT^\ast M\to \wedge^2\sT^\ast M\,,\quad \zg(j^1A(x))=\xd A(x)\,.$$
and
$$\bar L:\wedge^2\sT^\ast M\to\zW^m\,,\quad \bar L(F)=\frac{1}{2}F\wedge\star F\,,$$
where $\star$ is the `Hodge star' associated with the metric.

{Taking now $L=\bar L\circ\zg:\sJ^1\sT^\ast M\to\zW^m$ as our Lagrangian, we get
$$\xd^\sv L(j^1A)(j^1B)=\frac12(\xd B\wedge\star F+F\wedge\star\xd B)=\xd B\wedge\star F\,,$$
where $F=\xd A$.}
The Lagrangian is constant on fibres of the projection $\zg$. As the phase space
we get $\mathcal{P}\simeq \sT^\ast M\times_M W^\ast\otimes\Omega^m$. Since
$W^\ast$ can also be split into symmetric and antisymmetric part, we have
$\mathcal{P}\simeq \sT^\ast M\times_M (\vee^2\sT M\oplus\wedge^2\sT M)\otimes\Omega^m$.
The Legendre map $\lambda: \sJ^1 \sT^\ast M\rightarrow\mathcal{P}$ associated with the
electromagnetic Lagrangian reads
$$\lambda(\sj^1 A(x))=(\;A(x),\; 0,\; G(\xd A(x))\otimes\eta\;).$$
The symmetric part of the momentum vanishes as a consequence of the fact that the Lagrangian depends only on the
antisymmetric part of the jet. The only nontrivial part of the momentum is then a bivector density or (according to Weyl duality) an odd two-form. The constitutive set $\mathcal{D}_L$ on the Lagrangian
side is a subset of $\sJ^1\sT^\ast M\times\sJ^\ast\sT^\ast M\otimes\Omega^m$
given by
$$\mathcal{D}_L=\alpha^{-1}(\xdv L(\sJ^1\sT^\ast M))=\{(j^1A(x),\Phi(x)): \bar\za(\Phi(x))=\xd^\sv L(j^1A(x)))\}\,.$$
{A 1-form $A$ satisfies the Euler-Lagrange equation if $\xd^\sv L(j^1(A))$ is $\za$-related to the first jet $j^1(\chi)$ of a section $\chi=X^k\ot\zvy_k$ of $\cP$, i.e., for all 1-forms $B$,
$$\xd B\wedge\star F=\xd(B\wedge\star F)+B\wedge\xd\star F=\xd\left(\langle X^k,B\rangle\right)\zvy_k\,.$$
}
{It is easy to see, that $\xd(B\wedge\star F)$ is always of the required form, and $B\wedge\xd\star F$ is never, except for the case $\xd\star F=0$.}
{In this way we have obtained the Maxwell equations (without sources).}

\medskip
On the Hamiltonian side of the triple, we need a generating object of the constitutive set in the form of a section
of the bundle $\sJ^+\sT^\ast M\rightarrow \sT^\ast M\times_M W^\ast$ (or, more generally, a section supported on a submanifold or a family of sections). Out of the general theory we know that $\mathcal{D}_L$ is generated for sure by
the family of sections corresponding to the following family of density valued functions:
$$H:\sJ^+ \sT^\ast M\times_{\sT^\ast M}\sJ^1\sT^\ast M\rightarrow \Omega^m,\qquad
H(\varphi, \sj^1A)=\varphi(\sj^1 A)-L(\sj^1 A).$$
Critical points of this family are given by the Legendre map, i.e. $(\varphi,\sj^1 A)$ is critical if $\lambda(\sj^1 A)$ equals the linear part of $\varphi$. It follows that the generating family $H$  can be replaced by a simpler generating object, namely one section $h$ supported on the submanifold $\lambda(\sJ^1\sT^\ast M)$,
$$\lambda(\sJ^1\sT^\ast M)=\{(A, r,p)\in\sT^\ast M\times_M \vee^2\sT^\ast M\times_M\wedge^2\sT^\ast M:\quad
r=0\}.$$
The value $h$ at $(A, 0, p)$ is an affine map on the fibre of $\sJ^1\sT^\ast M$ over $A\in\sT^\ast M$. To know
$h(A,0,p)$ we have to know
how it acts on the jet $\sj^1\alpha(x)$, where $\alpha(x)=A$. For $h$, we get the following formula:
$$h(A, 0, p)(\sj^1\alpha(x))=\langle\; p,\; \xd\alpha(x)\;\rangle\omega-\frac12 p\wedge\star p\,.$$

\section*{Appendix: The proof of theorem \ref{th:1}}
\label{sec:17}

\begin{proof}
Let $U\subset V$ be a vector subbundle of $V$ such that $V\simeq W\oplus_N U$. As a set then, $V\simeq W\times_N U$. Once we have chosen
$U$, we can get the following identifications:
$$
V\simeq U\times_N W,\qquad  V^\ast\simeq U\times_N W^*, \qquad V^\dag_W\simeq (U\times_N W^*)\times\R,$$
and finally
\begin{equation}\label{eq:pv}
\sP V^\dag_W\simeq \sT^*(U\times_N W^\ast)\simeq \sT^*U\times_{\sT^*N} \sT^*W^*\,.
\end{equation}
On the other hand,
$$\sT^*V=\sT^*(U\times_N W)\simeq \sT^*U\times_{\sT^*N} \sT^*W
$$
and we obtain a symplectomorphism, using identity on the first factor and the canonical double vector bundle morphism $\mathcal{R}_W:\sT^\ast W\rightarrow\sT^\ast W^\ast$ (see (\ref{eq:izoR})) composed with
minus identity on the second factor.  Of course, the identifications  we have used depend on the choice of $U$. We have to show now that
the isomorphism between $\sP V^\dag$ and $\sT^*V$ is canonical, even if we pass through two
non-canonical maps.

The vector bundle $V$ and its subbundle $W$ give rise to the following canonical structures. We have the affine
bundle $\zt:V\rightarrow V\slash W$, the subbundle $W^0$ of $V^\ast$, the affine bundle $\pi:V^\ast\rightarrow V^\ast\slash W^0$
and  canonical isomorphisms $(V\slash W)^\ast\simeq W^0$, $W^\ast\simeq V^\ast\slash W^0$. The choice of
$U$ gives rise to two isomorphisms:
\begin{align*}
F:V\slash W &\longrightarrow U \subset V\,, \\
G:V^\ast\slash W^0\simeq W^\ast &\longrightarrow U^0 \subset V^\ast\,,
\end{align*}
where, clearly, $W^*=U^0, U^*=W^0\subset V^*$ are the annihilators of the subbundles $U,W\subset V$, respectively.
Choosing $U'$ instead of $U$, we get $F'$ and $G'$. Choosing an appropriate linear map $A:V\slash W\rightarrow W$, we can write
\begin{align*}
F'(q)&=F(q)+A(q)\,,\\
G'(a)&=G(a)-A^\ast(a)\,.
\end{align*}
For any $v\in\tau^{-1}(q)$ and $\alpha\in\pi^{-1}(a)$, we get two decompositions:
\begin{equation}\label{eq:decompositons}
v=w+F(q)=w'+F'(q)\,, \qquad
\alpha=G(a)+b=G'(a)+b'\,,
\end{equation}
with
\begin{equation}\label{eq:decompositons2}w'=w-A(q)\quad\text{and}\quad b'= b+A^\ast(a)\,.\end{equation}
Using $U$ and $U'$, we get also two decompositions:
$$V^\dag_W\;\simeq\; U\times_N U^0\times\R\;\simeq\; U'\times_N (U')^0\times\R.$$
As our considerations are local, we can assume that the bundles are trivial and ignore the basic coordinates of $N$. An element $\varphi\in V^\dag_W$ over $q$, with the linear part equal to $a$, is represented by
$(F(q), G(a), r)$ or $(F'(q), G'(a), r')$, where
\begin{equation}\label{eq:equal}r'=r+\langle G(a), A(q)\rangle.\end{equation}
A section $\sigma$ of the bundle $V^\dag_W\rightarrow V\slash W\times_N W^\ast$ in the neighbourhood
of a point $(q_0, a_0)\in V\slash W\times_N W^\ast$ can be written as
$$\sigma(q_0+\delta q, a_0+\delta a)=(F(q_0+\delta q), G(a_0+\delta a), r(q_0+\delta q,a_0+\delta a))\,.$$
For the purpose of studying $\sP V^\dag_W$, it is enough to consider sections that are affine with respect
to $\delta q$ and $\delta a$. Such sections (in the decomposition given by $U$) are defined by two elements
$w\in W$ and $b\in W^0$,
$$r(q_0+\delta q,a_0+\delta a)=\langle b, F(q_0+\delta q)\rangle-\langle G(a_0+\delta a), w\rangle\,.$$
For the pair $(v_0, \alpha_0)\in V\times V^* $
such that $\tau(v_0)=q_0$ and $\pi(\alpha_0)=a_0$, we get two decompositions as in (\ref{eq:decompositons}) and
(\ref{eq:decompositons2}). Now we should check whether the two sections $\sigma$ for $(w_0, b_0)$ and
$\sigma'$ for $(w_0', b_0')$ are equivalent, i.e. whether they have the same affine differential.
We have:
\begin{align*}
&\sigma(q_0+\delta q, a_0+\delta a)=(F(q_0+\delta q), G(a_0+\delta a), r(q_0+\delta q,a_0+\delta a))\,,\\
\intertext{with}
&r(q_0+\delta q,a_0+\delta a)=\langle b_0, F(q_0+\delta q)\rangle-\langle G(a_0+\delta a), w_0\rangle\,, \\
\intertext{and}
&\sigma'(q_0+\delta q, a_0+\delta a)=(F'(q_0+\delta q), G'(a_0+\delta a), r'(q_0+\delta q,a_0+\delta a))\,,\\
\intertext{with}
&r'(q_0+\delta q,a_0+\delta a)=\langle b'_0, F'(q_0+\delta q)\rangle-\langle G'(a_0+\delta a), w'_0\rangle\,.
\end{align*}
The difference between those two sections is a function on $V\slash W\times W^\ast$
(we have used (\ref{eq:equal}) here) which reads as
\begin{multline*}\sigma'(q_0+\delta q, a_0+\delta a)-\sigma(q_0+\delta q, a_0+\delta a)=\\
r'(q_0+\delta q,a_0+\delta a)-r(q_0+\delta q,a_0+\delta a)-\langle G(a_0+\delta a), A(q_0+\delta q)\rangle=\\
\langle G(a_0), A(q_0)\rangle-\langle G(\delta a),A(\delta q)\rangle.
\end{multline*}
The differential of this function at $(q_0, a_0)$ is equal to zero, since the first term is constant
and the second is quadratic. It means that the affine differentials $\underline{\xd}\sigma(q_0, a_0)$
and $\underline{\xd}\sigma'(q_0, a_0)$ are equal. The differential is therefore given by $(v_0, \alpha_0)$\,.
\end{proof}

\begin{remark} Note that a side result of the above theorem is the following `exotic' double vector-affine bundle structure on the bundle $\sP(V^\dag_W)\simeq\sT^\ast V$:
$$\xymatrix@C-10pt{
& \sP(V^\dag_W)\simeq\sT^\ast V\ar[dl]_{\tau\times\pi}\ar[dr]^{pr_1} & \\
V\slash W\times_N W^\ast & & V \,.\\
}$$
The isomorphisms $\sP(V^\dag_W)\simeq\sT^\ast V$, where $W$ runs through all vector subbundles of $V$, yields in particular $id_{\sT^\ast V}$ for $W=\{ 0\}$ and $-\mathcal{R}_V:\sT^\ast V\to\sT^\ast V^\ast$, for $W=V$ (see (\ref{eq:izoR})). In the theorem \ref{th:1}, the canonical symplectomorphism can be replaced by a canonical anti-symplectomorphism, its negative, which is often used in physical theories.
\end{remark}

\begin{remark} An alternative `symplectic' proof of theorem \ref{th:1} is also possible. As in the case of the canonical
isomorphism $\sT^\ast V\simeq\sT^\ast V^\ast$, there is a symplectic relation $S\subset\sT^\ast V\times\sT^\ast V^\dag_W$
generated by the evaluation $V^\dag_W\times_{V\slash W } V\ni(\varphi, v)\mapsto\varphi(v)\in \R$. The relation, composed with
symplectic reduction with respect to a certain coisotropic submanifold in $\sT^\ast V^\dag_W$, gives the isomorphism
between $\sT^\ast V$ and $\sP V^\dag_W$. It is clear from the construction that the isomorphism is a symplectomorphism
and a double vector-affine bundle morphism.
\end{remark}


\end{document}